\documentclass[useAMS,usenatbib]{mn2e}

\usepackage{amsmath,amssymb,mathrsfs,graphicx,float,latexsym,url}
\usepackage{epstopdf,ragged2e}
\usepackage{natbib}
\usepackage{hyperref}
\usepackage{varwidth}
\usepackage{supertabular}
\usepackage{subcaption}
\usepackage{multirow}
\usepackage[figuresright]{rotating}
\usepackage[usenames,dvipsnames]{color}
\usepackage{mathtools}
\usepackage{amsbsy}
\usepackage{bm}
\usepackage{float}
\usepackage{changes}

\def\apj{{ApJ}}

\def\apjs{{The Astrophysical Journal Supplement}}
\def\apjl{{ApJL}}
\def\bain{{BAN}}
\def\jcap{{JCAP}}
\def\pasp{{The Publications of the Astronomical Society of the Pacific}}
\def\aap{{A\&A}}

\def\baas{{Bulletin of the American Astronomical Society}}

\def\araa{{Annual Review of Astronomy and Astrophysics}}

\def\mnras{{MNRAS}}

\def\nat{{Nature}}
\def\na{{New Astronomy}}
\def\aapr{{The Astronomy and Astrophysics Review}}

\def\prd{{Physical Review D}}
\def\prl{{Phys. Rev. Lett.}}

\def\apss{{Ap\&SS}}
\def\04a{{2004 a}}
\def\04b{{2004 b}}

\def\msun{{\rm M}_{\odot}}
\def\rsun{{\rm R}_{\odot}}

\def\mdi{M_{\rm i,1}}
\def\mci{M_{\rm i,2}}
\def\ai{a_{\rm i}}
\def\ei{e_{\rm i}}
\def\mpy{{\rm M}_{\odot}\,{\rm yr}^{-1}}
\def\zsun{{\rm Z}_{\odot}}
\def\zm{z_{\rm m}}

\begin{document}

\title{On the Formation of GW190521-like Binary Black Hole Merger Systems
}
\author[Cui et al.]{Zhe Cui$^1$
and Xiang-Dong Li$^{1,2}$\thanks{Email: lixd@nju.edu.cn} \\
$^1$School of Astronomy and Space Science, Nanjing University, Nanjing 210023, China\\
$^2$Key Laboratory of Modern Astronomy and Astrophysics, Nanjing University, Ministry of Education, Nanjing 210023, China\\
}

\date{Accepted . Received ; in original form }

\pagerange{\pageref{firstpage}--\pageref{lastpage}} \pubyear{?}

\maketitle
\label{firstpage}

\begin{abstract}\label{sec:abs}

\noindent GW190521 is the most massive merging binary black hole (BBH) system detected so far. At least one of the component BHs was measured to lie within the pair-instability supernova (PISN) mass gap ($\sim 50-135\;\msun$), making its formation a mystery. However, the transient observed signal allows alternative posterior distributions. There was suggestion that GW190521 could be an intermediate-mass ratio inspiral (IMRI), with the component masses  $m_1\sim 170\;\msun$ and $m_2\sim 16 \;\msun$, happening to straddle the PISN mass gap. Under this framework, we perform binary population synthesis to explore the formation of GW190521-like systems via isolated binary evolution.
We numerically calculate the binding energy parameter for massive stars at different metallicities, and employ them in our calculation for common envelope evolution.
Our results prefer that the progenitor binaries formed in metal-poor environment with $\rm Z\leq0.0016$. The predicted merger rate density within redshift $z=1.1$ is $\sim 4\times 10^{-5}-5\times 10^{-2} \,\rm Gpc^{-3}yr^{-1}$. We expect that such events are potentially observable by upcoming both space and ground-based gravitational wave detectors.
\end{abstract}



\section{Introduction}
\label{sec:intro}

Detection of gravitational waves (GWs) serves us an alternative way to observe the universe. Since the first GW event GW150914 was discovered by the ground-based detectors advanced LIGO (aLIGO) and later joined advanced Virgo, the number of binary black hole (BBH) merger events has increased to $\sim$100 \citep{Abbott2016PhRvL.116f1102A, Abbott2016PhRvX...6d1015A, Abbott2019PhRvX...9c1040A, Abbott2021PhRvX..11b1053A, Abbott2021arXiv211103606T}. The observed GW signals are classified as coalescing BBH, binary neutron star (BNS) and neutron star-black hole (NSBH) systems. GW170817 is the only GW source with electromagnetic (EM) counterpart definitely observed \citep{Abbott2017ApJ...848L..13A}.

GW190521, observed on May 21, 2019 at 03:02:29 UTC, is the most massive merging BBH system detected so far \citep{Abbott2020PhRvL.125j1102A, Abbott2020ApJ...900L..13A}. The association of GW190521 with the candidate counterpart ZTF19abanrhr reported by Zwicky transient facility (ZTF) \citep{Graham2020PhRvL.124y1102G} is still inconclusive \citep{Ashton2021CQGra..38w5004A, Nitz2021ApJ...907L...9N, Palmese2021ApJ...914L..34P}. Under the assumption that GW190521 is a quasi-circular BBH coalescence, the estimated individual component masses are $m_1=85_{{-14}}^{+21}\msun$, $m_2=66_{{-18}}^{+17}\msun$, and the total mass $150_{{-17}}^{+29}$ $\msun$ within 90\% credible region, providing direct evidence of intermediate mass BHs (IMBHs) \citep{Abbott2020PhRvL.125j1102A, Abbott2020ApJ...900L..13A}. \cite{Gamba2021arXiv210605575G} drew similar results but under hyperbolic orbit hypothesis. \cite{Romero-Shaw2020ApJ...903L...5R} claimed that GW190521 may be an eccentric bianry merger with aligned spins. \cite{Gayathri2022NatAs...6..344G} interpreted this signal under the combination of both eccentricity and spin precession configuration. \cite{Barrera2022ApJ...929L...1B} estimated the ancestral mass of GW190521 and also favored the heaviest parental BH mass in the pair-instability supernova (PISN) mass gap \citep[between $\sim 50-135 \msun$,][]{Yusof2013MNRAS.433.1114Y,Belczynski2016A&A...594A..97B}.

Since the BH masses in GW190521-like events challenge the standard stellar evolutionary theory, there have been various models proposed to interpret their formation, including dynamical binary formation in dense stellar clusters \citep{Rodriguez2019PhRvD.100d3027R, Romero-Shaw2020ApJ...903L...5R, Fragione2020ApJ...902L..26F, Anagnostou2020arXiv201006161A, Gamba2021arXiv210605575G, Arca-Sedda2021ApJ...920..128A, Rizzuto2022MNRAS.512..884R}, additional gas accretion and hierarchical mergers in active galactic nuclei (AGNs) \citep[][and references therein]{Tagawa2020ApJ...898...25T,Tagawa2021ApJ...908..194T}, and the primordial BH scenarios \citep{De Luca2021PhRvL.126e1101D}. Alternatively, \cite{Palmese2021PhRvL.126r1103P} suggested that GW190521 may be the merger of central BHs from two ultradwarf galaxies. However, the origin of this event as an isolated binary still cannot be excluded \citep{ Belczynski2020ApJ...905L..15B,Farrell2021MNRAS.502L..40F, Kinugawa2021MNRAS.501L..49K, Tanikawa2021MNRAS.505.2170T}. In addition, the exact boundaries of the PISN mass gap are in dispute, due to the uncertainties in stellar evolution and SN simulation, which may entail a reassessment \citep{Woosley2017ApJ...836..244W, Marchant2019ApJ...882...36M, Farmer2019ApJ...887...53F, Mapelli2020ApJ...888...76M, Vink2021MNRAS.504..146V}.

Nevertheless, GW190521 is qualitatively different from previous GW sources, not only because it was the most massive GW source observed to date, but also this transient signal was found with only a short duration of approximately 0.1 s, and only around four cycles in the frequency band $30-80$ Hz, so multimodal posterior distributions would be consequently ineluctable \citep{Fishbach2020ApJ...904L..26F, Nitz2021ApJ...907L...9N, Bustillo2021PhRvL.126t1101B, Estells2022ApJ...924...79E}. Among them,  \cite{Nitz2021ApJ...907L...9N} suggested that GW190521 may be an intermediate-mass-ratio inspiral (IMRI), with the component masses of $m_1\sim 170\,\msun$ and $m_2\sim 16\,\msun$, straddling the PISN mass gap. Comparison of the parameters derived by \cite{Abbott2020PhRvL.125j1102A} and \cite{Nitz2021ApJ...907L...9N} are shown in Table \ref{table1}.

Inspired by the results of \cite{Nitz2021ApJ...907L...9N}, here we attempt to interpret the formation of GW190521 assuming that it was an IMRI through isolated binary evolution, and investigate the properties of their progenitor binaries as well as the possible distributions of natal kicks on the two component BHs, which had promoted their coalescence within the Hubble time $\tau_{\rm H}$. The information on the BH kicks is crucial in understanding the formation of massive BHs.

The paper is structured as follows. In section \ref{sec:calc} we describe the main features of our binary population synthesis (BPS) models. The calculated results of BPS are   presented in section \ref{sec:results}. We then discuss our results in section \ref{sec:dis}, and summarize our main conclusions in section \ref{sec:summary}.

\begin{table*}
	\caption{The derived primary BH mass $m_1$, secondary BH mass $m_2$, total mass $M_{\rm tot}$, dimensionless spin parameters of individual BH and effective spin parameter $\protect\overrightarrow{\chi_1}$, $\protect\overrightarrow{\chi_2}$, and $\protect\chi_{\rm eff}$ for GW190521 in the source frame. Data are cited from \protect\cite[][A20]{Abbott2020PhRvL.125j1102A} and \protect\cite[][NC21]{Nitz2021ApJ...907L...9N}, respectively. In the latter work, $Prior_{q^{*}-M}$ denotes the prior uniform in mass ratio and total mass, $Prior_{m_{1,2}}$ the prior uniform in component mass ($m_{1,2}$), respectively. Each value is within the 90\% credible interval. Note here $q^{*}$ is the ratio of the larger mass to the smaller mass.}
	\centering
\renewcommand\arraystretch{1.5}
		\begin{tabular*}{0.83\textwidth}{cc|cccccc}
			\hline\hline
			\multicolumn{2}{c}{Model} & $m_1(\msun)$ & $m_2(\msun)$ & $M_{\rm tot}(\msun)$& $|\overrightarrow{\chi_1}|$& $|\overrightarrow{\chi_2}|$& $\chi_{\rm eff}$\\
			\hline
			\multicolumn{2}{c|}{A20} & $85_{{-14}}^{+21}$ & $66_{{-18}}^{+17}$ & $150_{{-17}}^{+29}$& $0.69_{-0.62}^{+0.27}$& $0.73_{-0.64}^{+0.24}$ &  $0.08_{-0.36}^{+0.27}$ \\ \hline
			& $Prior_{q-M}$ & $168_{-61}^{+15}$ & $16_{-3}^{+33}$ & $184_{-30}^{+15}$ & $0.85_{-0.25}^{+0.11}$& -&  $-0.51_{-0.11}^{+0.24}$ \\ \cline{2-8}
			\ \ \ NC21\ \ \  & $Prior_{m_{1,2}}$ ($q^{*}<4$)  &  $100_{-18}^{+17}$ & $57_{-16}^{+17}$ & $156_{-15}^{+21}$ &  $0.72_{-0.59}^{+0.25}$& -&  $-0.16_{-0.40}^{+0.42}$\\ \cline{2-8}
			& $Prior_{m_{1,2}}$ ($q^{*}>4$)&$166_{-35}^{+16}$ &$16_{-3}^{+14}$&$183_{-27}^{+15}$ &  $0.87_{-0.16}^{+0.10}$&- & $-0.53_{-0.12}^{+0.14}$ \\
			\hline\hline
		\end{tabular*}
		\label{table1}
\end{table*}

\section{Model} \label{sec:calc}
All calculations are carried out by using the BPS code {\tt\string BSE}, originally developed by \citet{Hurley2000MNRAS.315..543H,Hurley2002MNRAS.329..897H} and its updated version {\tt\string BSEEMP}\footnote{https://github.com/atrtnkw/bseemp.} \citep{Tanikawa2020MNRAS.495.4170T, Tanikawa2021ApJ...910...30T}, with the extension to very massive ( up to $M\sim 1300\,\msun$) and extremely metal-poor stars (down to $Z=10^{-8}\zsun$), based on the stellar models computed by the HOSHI code. The {\tt\string BSEEMP} code also takes advantage of new stellar-wind and remnant-formation prescriptions, as well as the implementation of pair-instability and pulsational-pair-instability supernova (PISN/PPISN), which all play a role in stellar/binary evolution.

\subsection{Stellar wind mass loss} \label{sub:wind}
The masses of BHs are predominantly set by their presupernova masses, which are mainly affected by stellar wind mass loss and binary interactions. Mass loss via stellar winds can significantly influence the fate of massive stars \citep{Fryer2002ApJ...578..335F}. Here we use the semi-empirical stellar wind prescription (referred as {\tt\string Vink et al.\,winds}) in \cite{Belczynski2010ApJ...714.1217B}, and consider metallicity dependence for Luminous Blue Variables (LBVs) \citep{Tanikawa2021ApJ...910...30T}. It has been demonstrated that this wind prescription results in more massive pre-supernova objects and heavier BHs compared with the traditional one \citep{Belczynski2010ApJ...714.1217B}. We ignore  the influence of stellar rotation on wind loss.

For massive O and B stars, the wind mass loss rate $\dot{M}_{\rm W}$ in units of $\mpy$ is \citep{Vink2001A&A...369..574V}
\begin{equation}
	\begin{array}{ll}
		\log(\dot{M}_{\rm W,OB}) = & -6.688+2.210 \log(L/10^{5}) \\
		& -1.339 \log(M/30) -1.601 \log(V/2.0) \\
		& +0.85 \log(Z/\zsun) +1.07 \log(T/20000),
	\end{array}
	\label{eq:cw_ob}
\end{equation}
with $12500\,{\rm K} \leq T \leq 25000$\,K. Here $L$ and $M$ are the luminosity and the mass in Solar units respectively, $Z$ is metallicity, $T$ is the effective temperature of the star, and $V=v_\infty/v_{\rm esc}=1.3$ is the ratio of the wind velocity at infinity to the escape velocity from the star.

For hotter stars with $25000\,{\rm K} \leq T \leq 50000$ K,
\begin{equation}
	\begin{array}{ll}
		\log(\dot{M}_{\rm W,OB}) = & -6.697+2.194 \log(L/10^{5}) \\
		& -1.313 \log(M/30) -1.226 \log(V/2.0) \\
		& +0.85 \log(Z/\zsun)+0.933 \log(T/40000) \\
		& -10.92 [\log(T/40000)]^2
	\end{array}
	\label{eq:hw_ob}
\end{equation}
with $V=2.6$.

For LBVs beyond the \citeauthor*{Humphreys1994PASP..106.1025H} limit ($L>6 \times 10^5$ and $10^{-5} R L^{0.5}>1.0$, where $R$ is the stellar radius in solar units),
\begin{equation}
	\dot{M}_{\rm W,LBV}= f_{\rm LBV} \times 10^{-4}(Z/\zsun)^{0.86} \mpy,
	\label{eq:w_lbv}
\end{equation}
where $f_{\rm LBV}=1.5$ is a calibration factor. Obviously, lower $f_{\rm lbv}$ results in weaker LBV wind, leaving a heavier remnant \citep{Belczynski2010ApJ...714.1217B}.

The reduced Wolf-Rayet star mass loss with small H-envelope mass takes the form of metallicity-dependent power law,
\begin{equation}
	\dot{M}_{\rm W,WR} =  10^{-13} L^{1.5} \left({\frac{Z}{\zsun}}\right)^{m}(1.0-\mu) \mpy,
	\label{eq:w_wr}
\end{equation}
with
	 	\[\mu = \left({\frac{M-M_{\rm He}}{M}}\right)\min\left\{{5.0,\max[1.2,(\frac{L}{7\times10^4})^{-0.5}]}\right\},\]
$m=0.86$ describing the dependence of wind mass loss on metallicity, and $M_{\rm He}$ the He core mass of the star \citep{Vink2005A&A...442..587V}.

For other stars, we use the wind prescriptions described in \cite{Hurley2000MNRAS.315..543H}.

\subsection{Black hole formation}\label{sub:bh formation}
Stars are powered by burning their core fuels to heavier elements step by step. For massive stars, this process continues until an iron core is built up in the stellar center. As the fusion of iron does not produce further energy, burning halts. Later, stars contract on their own weights, leading to accelerating processes of electron capture and core element dissociation. These processes dramatically reduce the pressure that should have resisted their self gravity, triggering an runaway core-collapse process. Collapse halted by nuclear forces and neutron degeneracy pressure and form a proto-NS. Explosion launches after the ``bounce" of the core, and part or all of the expelled stellar envelope will fall back and accrete onto the proto-NS, which may eventually collapse into a BH \citep{Fryer2012ApJ...749...91F}.

There are still many uncertainties associated with the physics of the SN mechanism. Here we use the delayed SN prescription of \cite{Fryer2012ApJ...749...91F} (hereafter F12-delayed), where the explosion did not lunch until over $\sim 250$ ms after the collapse. For a massive star with pre-SN mass of $M_{\rm SN}$ and CO core mass of $M_{\rm CO}$, the expected BH remnant mass $M_{\rm BH}$ is estimated as follows,
\begin{equation}
	M_{\rm BH} = 0.99M_{\rm BH, bar}=0.99(M_{\rm proto}+M_{\rm fb}),
	\label{eq:m_bh}
\end{equation}
where $M_{\rm BH, bar}$ is the baryonic mass\footnote{The baryonic mass is reduced by the neutrinos that are lost \citep{Burrows1986ApJ...307..178B}, and we assume that for BHs the gravitational mass $M_{\rm BH}$ is $99\%$ of the baryonic mass $M_{\rm BH,bar}$ for our considered massive BBHs.}, $M_{\rm proto}$ is the proto-NS mass after core collapse,
\begin{equation}
	M_{\rm proto} =
	\left\{ \begin{array}{rll}
		1.2 \msun & M_{\rm CO} < 3.5 \msun\\
		1.3 \msun & 3.5 \leq M_{\rm CO} < 6.0 \msun \\
		1.4 \msun & 6.0 \leq M_{\rm CO} < 11.0 \msun \\
		1.6 \msun & M_{\rm CO} \geq 11.0 \msun,
	\end{array}
	\right.
	\label{eq:m_proto}
\end{equation}
and $M_{\rm fb}$ is the amount of material falls back to the proto-NS,
\begin{equation}
	M_{\rm fb} =
	\left\{ \begin{array}{lll}
		0.2 \msun & M_{\rm CO} < 2.5 \msun\\
		0.5 M_{\rm CO}-1.05 \msun & 2.5 \leq M_{\rm CO} < 3.5 \msun \\
		(f_1 M_{\rm CO} + f_2)(M_{\rm SN}-M_{\rm proto})  & 3.5 \leq M_{\rm CO} < 11.0 \msun\\
		M_{\rm SN}-M_{\rm proto} & M_{\rm CO} \geq 11.0 \msun,
	\end{array}
	\right.
	\label{eq:m_fb}
\end{equation}
where $f_1=0.133 - {0.093 \over M-M_{\rm proto}}$, and $f_2=-11 f_1 + 1$.
We define $f_{\rm fb}$ = $M_{\rm fb} /(M_{\rm SN}-M_{\rm proto})$ as the fallback fraction during the BH formation, which is important in determining the BH's natal kick in some kick prescriptions.

Stars with He core mass $M_{\rm He}$ in the range of $\sim 35-60\,\msun$ are subjected to PPISNe \citep{Heger2002ApJ...567..532H, Yusof2013MNRAS.433.1114Y, Belczynski2016A&A...594A..97B, Marchant2019ApJ...882...36M, Stevenson2019ApJ...882..121S, Leung2019ApJ...887...72L}, with most of the mass above the core stripped by a set of pulsations, leaving behind the BHs with mass significantly smaller than they would be if only accounting for the core-collapse SNe. We adopt the prescription of PPISNe in \citet{Marchant2019ApJ...882...36M}, who computed an array of H-free metal-poor ($0.1\zsun$) single-star models based on the standard $^{12}C(\alpha,\gamma)O^{16} $ reaction rate to evaluate the PPISN mass loss. And the BH masses after PPISNe can be estimated as:
\begin{equation}
	M_{\rm BH} = M_{\rm He}\sum_{i=0}^{7}\zeta_i(\frac{M_{\rm He}}{\msun})^i,
	\label{eq:ppisn}
\end{equation}
where $\zeta_i$ are the polynomial fitting coefficients of \cite{Marchant2019ApJ...882...36M}'s PPISN prescription given by \cite{Stevenson2019ApJ...882..121S} (as listed in Table \ref{table2}). Note that the remnant mass is a non-monotonic function of the initial stellar mass.
\begin{table}
	\caption{Coefficients in Equation~\ref{eq:ppisn}}
	\centering
	\renewcommand\arraystretch{1.5}
	\begin{tabular}{c|c}
		\hline\hline
		Coefficient & Value  \\ \hline
		$\zeta_0$   &  7.39643451 $\times 10^{3}$  \\
		$\zeta_1$   & -1.13694590 $\times 10^{3}$  \\
		$\zeta_2$   &  7.45060098 $\times 10^{1}$  \\
		$\zeta_3$   & -2.69801221 $\times 10^{0}$ \\
		$\zeta_4$   & 5.83107626 $\times 10^{-2}$  \\
		$\zeta_5$   & -7.52206933 $\times 10^{-4}$  \\
		$\zeta_6$   & 5.36316755 $\times 10^{-6}$  \\
		$\zeta_7$   & -1.63057326 $\times 10^{-8}$  \\
		\hline\hline
	\end{tabular}
	
	\label{table2}
\end{table}

More massive stars with $60\,\msun\leq M_{\rm He} \leq 135\,\msun$ are subjected to PISNe, the entire star is completely disrupted with no remnant left\footnote{\cite{Belczynski2020ApJ...905L..15B} recently suggested that if the $ ^{12}C(\alpha,\gamma)O^{16} $ reaction rate is $ 3\sigma $ lower than its standard rate, star with Helium core mass $ M_{\rm He} \sim 90 \msun$ can avoid PISN and evolve to a mass gap BH.}. Stars with $M_{\rm He}>135\,\msun$ are assumed to directly collapse to BHs.

\subsection{Supernova kicks}\label{sub:kick}
As BH natal kicks suffer from a lack of stringent constraints from both observation and theory \citep{Willems2005ApJ...625..324W, Fragos2009ApJ...697.1057F, Repetto2012MNRAS.425.2799R, Repetto2017MNRAS.467..298R,  Repetto2015MNRAS.453.3341R, Mandel2016MNRAS.456..578M, Belczynski2016ApJ...819..108B}, we adopt three different natal kick prescriptions ($kick_{\rm F}$) as follows \citep{Banerjee2020A&A...639A..41B}:
\begin{itemize}
	\item [(1)]  Standard fallback-controlled kick (hereafter $ k1 $)\\
	The BH natal kick velocities $v_{\rm kick, BH}$ are scaled linearly with the NS natal kick velocities $v_{\rm kick, NS}$ by a factor $(1-f_{\rm fb})$ \citep{Fryer2012ApJ...749...91F, Giacobbo2018MNRAS.474.2959G},
	\begin{equation}
		v_{\rm kick,BH} = v_{\rm kick,NS}(1-f_{\rm fb}).
		\label{eq:k1}
	\end{equation}
	\item [(2)] Convection-asymmetry-driven natal kick (hereafter $ k2 $)\\
	The BH natal kicks are produced by the convection asymmetries of the collapsing SN core \citep{Scheck2004PhRvL..92a1103S,Fryer2006ApJS..163..335F}, so
	\begin{equation}
		v_{\rm kick,BH} =
		\left\{ \begin{array}{ll}
			v_{\rm kick,NS}\frac{<M_{\rm NS}>}{M_{\rm BH}}(1-f_{\rm fb})  & {\rm if}\ M_{\rm CO} \leq 3.5 \msun,\\
			k_{\rm conv}v_{\rm kick,NS}\frac{<M_{\rm NS}>}{M_{\rm BH}}(1-f_{\rm fb})  & {\rm if}\ M_{\rm CO} > 3.5 \msun.\\
		\end{array}
		\right.
		\label{eq:k2}
	\end{equation}
	In this equation, $k_{\rm conv}$ is an efficiency factor (somewhere between 2 and 10, and we set $k_{\rm conv}$ = 5 here), and $<M_{\rm NS}>$ is a typical NS mass, taken to be $1.4\,\msun$.
	\item [(3)] Neutrino-driven natal kick (hereafter $ k3 $)\\
	The BH natal kicks are produced through asymmetric neutrino emission \citep{Fuller2003PhRvD..68j3002F,Fryer2006ApJS..163..335F},
	\begin{equation}
		v_{\rm kick,BH} = v_{\rm kick,NS}\frac{ \min(M_{\rm eff},M_{\rm BH})}{M_{\rm BH}},
		\label{eq:k3}
	\end{equation}
\end{itemize}
where $ M_{\rm eff} $ (usually between  $ 5\,\msun $ and  $ 10\,\msun $) is the effective remnant mass, and we let $ M_{\rm eff} = 7\,\msun $ \citep{Banerjee2020A&A...639A..41B}.

To constrain the allowed velocity range in different kick prescriptions, we take a flat distribution of $v_{\rm kick,NS}$ in the range of $0-1000\rm \,kms^{-1}$, and assume that the supernova kicks are isotropically distributed and the mass is instantaneously lost at the moment of SN. Then $v_{\rm kick,BH}$ can be obtained from the equations mentioned above for different kick prescriptions. When we calculate the merger rate density we adopt a more realistic, predetermined Maxwellian distribution for the NS kick velocity.

Note also that in both $ k1 $ and $ k2 $ prescriptions there is no natal kick for BHs formed through direct core collapse ($ f_{\rm fb} $ = 1.0).

\subsection{Natal BH spins}\label{sub:spin}
The BH spins are modeled following \cite{Tanikawa2021ApJ...910...30T}. We assume zero natal spin of the zero-age main sequence (ZAMS) star, which then evolves due to stellar evolution, stellar winds, and binary interactions. Newborn BHs inherit their progenitor's spin angular momenta, except for the PPISN events where zero BH spin parameters are assumed. If the spin angular momenta of the BH progenitors are larger than those of extreme Kerr BHs, the BH spin parameters are forced to be unity. Thus the BH spin parameter can be expressed as:

\begin{equation}
	\overrightarrow{\chi} =
	\left\{\begin{array}{ll} 0 & \rm PPISN,\\
		\min(\frac{c}{GM^2}|\overrightarrow{S}|,1)\frac{\overrightarrow{L}}{|\overrightarrow{L}|} & \rm otherwise,\\
		\end{array}
		\right.
\end{equation}\label{chi}
where $\overrightarrow{S}$ and $M$ are the spin angular momentum and the mass of the BH progenitor just before its collapse respectively, $c$ the speed of light, $G$ the gravitational constant, and $\overrightarrow{L}$ the binary ortbial angular momentum.

The BH natal kicks would tilt $\overrightarrow{\chi}$ from $\overrightarrow{L}$. We choose the coordinate in which the $z$-axis is parallel to the orbital angular momentum vector just before the second BH formed, that is, the normalized orbital angular momentum vector is $(0,0,1)$. Then the normalized spin vectors of the first and second formed BH $\overrightarrow{\chi_1}$ and $\overrightarrow{\chi_2}$ can be written as:
\begin{equation}
\begin{array}{ll}
	\overrightarrow{\chi_1} =  (\sin\theta_1^{\prime}\cos\phi_1^{\prime},\sin\theta_1^{\prime}\sin\phi_1^{\prime},\cos\theta_1^{\prime}),\\
	\overrightarrow{\chi_2} =  (0,0,1),\\
\end{array}	
\end{equation}\label{chi_2}
where $\theta_1^{\prime}$ is the angle between the first BH spin vector and binary orbital angular momentum vector just before the second BH forms, $\phi_1^{\prime}$ is randomly chosen between $0$ and $2\pi$.
Finally the angles $\theta_1$ ($\theta_2$) between the first (second) formed BH spin and the final BBH orbital angular momentum vetoer just after the second BH formation can be expressed as:
\begin{equation}
\begin{array}{cc}
	\cos\theta_1 = \overrightarrow{\chi_1} \cdot \frac{\overrightarrow{L}}{|\overrightarrow{L}|},\\
	\cos\theta_2 = \overrightarrow{\chi_2} \cdot \frac{\overrightarrow{L}}{|\overrightarrow{L}|}.\\
\end{array}
\end{equation}\label{angle}
And we do not consider possible BH spin aligment with orbital angular momentum due to tides or mass transfer. The effective spin parameter $\chi_{\rm eff}$ of merging BBHs which reflects the spin-orbit alignment is defiend as:
\begin{equation}
\chi_{\rm eff}\equiv \frac{m_1 |\overrightarrow{\chi_1}|\cos\theta_1+m_2|\overrightarrow{\chi_2}|\cos\theta_2}{m_1+m_2},
\end{equation}\label{chi_eff}
where $m_1$ and $m_2$ are the merging BBH masses.

\subsection{Common-envelope evolution}\label{sub:CE}
For semi-detached binaries, a common-envelope (CE) phase occurs when the mass transfer becomes dynamically unstable or when the two stars (with at least one of them being a giant-like star) collide at orbital periastron before Roche lobe overflow (RLOF) \citep{Hurley2002MNRAS.329..897H}.
The CE phase plays a fundamental role in the formation of GW190521-like systems. Due to the extreme initial mass ratio $q\sim 0.1$, mass transfer from the massive primary star to the secondary star is usually dynamically unstable and leads to CE evolution. In addition, systems with large orbital eccentricities are likely to collide at periastron.
Binaries can survive the CE phase if the accretor's orbital energy is large enough to unbind the stellar envelope\footnote{See \cite{Hurley2002MNRAS.329..897H} and \cite{Tanikawa2022ApJ...926...83T} for a more comprehensive description of {\tt\string BSEEMP} prescriptions of mass transfer during RLOF and CE.}. Here we adopt the $ \alpha_{\rm CE}\lambda $ formalism \citep{de Kool1990ApJ...358..189D}, which can be expressed as:
\begin{equation}\label{eq:ebind-alpha}
	E_{\rm bind} = \alpha_{\rm CE}\biggl(-\frac{{\rm G}M_{\rm c}M_2}{2a_{\rm f,CE}}+\frac{{\rm G} M_1M_2}{2a_{\rm i,CE}}\biggr),
\end{equation}
where the envelop's binding energy,
\begin{equation}\label{eq:ebind-lambda}
	E_{\rm bind} = \int_{M_c}^{M_{\rm 1}}(-\frac{{\rm G} M(r)}{r}+\alpha_{\rm th}U)dm = -\frac{{\rm G}M_1M_{\rm env}}{\lambda R_{\rm RL}}.
\end{equation}
Here $\alpha_{\rm CE}$ is the efficiency of converting the released orbital energy to eject CE, $ \lambda $ the binding energy parameter depending on envelope's structure \cite[see][for details]{Ivanova2013A&ARv..21...59I}, $M_1$, $M_{\rm c}$ and $M_{\rm env}$ = ($M_1-M_{\rm c}$ ) the masses of the donor, donor's core and envelope respectively, $M_{2}$ the mass of the accretor, $a_{\rm i,CE}$ and $a_{\rm f,CE}$ the binary separation before and after the CE phase respectively, $R_{\rm RL}$ the donor's RL radius at the onset of the CE phase, $U$ the specific internal energy (including both thermal and recombination energies) of envelope, and $\alpha_{\rm th}$ the efficiency with which thermal energy can be used to eject the envelope. In this work $\alpha_{\rm th}$ = 1 is assumed.

Following \cite{Xu2010ApJ...716..114X} and \cite{Wang2016RAA....16..126W}, we calculate the binding energy parameter $\lambda$ for stars more massive than $60\,\msun$ with metallicity $Z=0.02,0.001,0.0001$ using the stellar evolution code MESA \citep[version 11701,][]{Paxton2011ApJS..192....3P, Paxton2015ApJS..220...15P, Paxton2018ApJS..234...34P, Paxton2019ApJS..243...10P}. We then calculate $\lambda_{\rm b}$ and $\lambda_{\rm g}$ with $E_{\rm bind}$ including and excluding the internal energy term $U$, respectively. Detailed models and fitting results of $\lambda$ are presented in Appendix \ref{app}. We only use $\lambda_{\rm b}$ in our following population synthesis calculations.

In order to explore the dependence of our results on $ \alpha_{\rm CE} $, we have run a set of simulations with $\alpha_{\rm{CE}} =  0.5$, 1.0, and 3.0 \citep[$\alpha_{\rm CE}> 1.0$ would occur when additional energy or angular momentum depositing into the giant's envelope is considered, see e.g.,][]{Soker2004NewA....9..399S}.

\subsection{Population synthesis}\label{sub:bps}
We perform BPS simulations of binary stars with the {\tt\string BSEEMP} code.
We assume that all the stars are in binaries, and exclude binary systems with at least one of the two components fills its RL at the beginning of evolution. The initial masses $\mdi$ of the primary stars are distributed in the range of $[300:900]\,\msun$ following the \cite{Kroupa2001MNRAS.322..231K} law. For the masses of the secondary stars $ \mci = q \mdi $, we adopt a flat distribution of the mass ratio $ q $ \citep{Sana2012Sci...337..444S}, and limit $\mci= [10:60]\,\msun$. We consider various metallicities with $Z = $ 0.0002, 0.0004, 0.0008, 0.0016, 0.0032, 0.0063, 0.0126 and 0.02. The initial orbital semi-major axis $\ai$ is assumed to be distributed uniformly in log-space and restricted to $[3:10^{7}]\,\rsun$. We consider both initially circular (hereafter `$\ei=0$' model) and eccentric (hereafter `$\ei=0\sim1$' model) orbit configurations in each case. In the latter model, the eccentricity follows a uniform distribution between 0 and 1.

We focus only on GW190521-like systems with the primary and secondary BH masses $m_1 = [150:180]\,\msun$ and $m_2 = [10:20]\,\msun$, so it is reasonable to only simulate binaries with a limited initial parameter ranges, and make sure that the simulated population parameters are complete to form GW190521-like systems.

Incorporating the three kick prescriptions ($kick_{\rm F}$ = $ k1, k2$ and $k3$) and three values of $\alpha_{\rm CE}=$ (0.5, 1.0 and 3.0), we perform 18 sets of BPS simulations of $ 10^7$ primordial binaries at each metallicity. We list the initial parameters of models with $Z \leq$ 0.0016 in Table \ref{table3}. For higher metallicity, we found that there is no GW190521-like system formed in our simulation.

\begin{table}
	\begin{center}
		\renewcommand\arraystretch{1.5}
		\caption{The initial parameters of our BPS models for both `$\ei=0$' and `$\ei=0\sim1$' models. Here $ \mdi $ and $ \mci$ are the initial primary and secondary masses respectively, and $ \ai $ is the initial orbital semi-major axis, all in Solar units. Because there is no BH as massive as $ \sim 150\,\msun$ formed in our calculation at $Z \geq 0.0032$, owing to the significant wind mass loss and PISN under our evolutionary assumptions, we only display the runs of $Z$ = 0.0002, 0.0004, 0.0008, and 0.0016 which can form BHs with mass in the range of interest.}
		\begin{tabular}{ccccccccc}
			\hline\hline
			$Z$  &\multicolumn{2}{c}{$0.0002$} & \multicolumn{2}{c}{$0.0004$} & \multicolumn{2}{c}{$0.0008$} & \multicolumn{2}{c}{$0.0016$}\\\hline
			$ \mdi\,[\msun]$ & \multicolumn{2}{c}{$300-450$} & \multicolumn{2}{c}{$350-500$} & \multicolumn{2}{c}{$400-700$} & \multicolumn{2}{c}{$500-900$}\\\hline
			$ \mci\,[\msun]$ & \multicolumn{8}{c}{$20-60$} \\\hline
			$\ai\,[\rsun]$   & \multicolumn{8}{c}{$ 10^{3.5}-10^7 $} \\\hline
			$kick_{\rm F}$ & \multicolumn{8}{c}{$ k1,\;k2,\;k3 $} \\\hline
			$ \alpha_{\rm CE}$ &\multicolumn{8}{c}{0.5, 1.0, 3.0}\\
			\hline\hline
		\end{tabular}
		\label{table3}
	\end{center}
\end{table}

We assume a nonconservative mass transfer prescription with the accretion efficiency being 0.5. We follow the evolution of the primordial binaries until the formation of BBHs. For a newborn BBH system with component masses $ m_1$ and $ m_2 $, orbital semi-major axis $ a_0 $ and eccentricity $ e_0 $, the inspiral time delay $ t_{\rm inspiral}(a_0,e_0) $, namely the time elapsed between the birth and the merger of the BBH, can be calculated \citep{Peters1964PhRv..136.1224P}:
\begin{equation}\label{eq:t_inspiral}
	t_{ \rm inspiral}(a_0,e_0) = \frac{12}{19} \frac{c_0^{4}}{\beta}
	\int_{0}^{e_0} \frac{e^{29/19} [1+(121/304)e^2]^{1181/2299}}{(1-e^2)^{3/2}}de,
\end{equation}
where
\begin{equation}\label{eq:c0}	
	c_0=a_0\ e_0^{-12/19}(1-e_0^2)\biggl(1+\frac{121}{304}e_0^2\biggr)^{-870/2299},	
\end{equation}
and
\begin{equation}\label{eq:beta}
	\beta=\frac{64}{5}\frac{G^3m_1m_2(m_1+m_2)}{c^5}.
\end{equation}

\subsection{Merger rate density }\label{sub:rate}
We then follow the procedure of \cite{Giacobbo2018MNRAS.480.2011G} to estimate the cumulative merger rate density $\mathcal{R}(z\leq z_{\rm det})$ of GW190521-like BBH systems within a given redshift $z_{\rm det}$,
\begin{equation}\label{eq:rate}
	\begin{aligned}
	\mathcal{R}(z \leq z_{\rm det}) = &\sum_{z=0.1}^{z=z_{\rm det}}(\frac{1}{t_{\rm lb}(z)-t_{\rm lb}(z -\Delta z)} \\
	& \sum _{z=15} ^{z=z_{\rm det}} (\frac{f_{\rm bin}}{2}
	\frac{{\mathcal{SFR}}(z)W_{\rm b}}{M_{*}} )[t_{\rm lb}(z+\Delta z) - t_{\rm lb}(z)]),\\
	\end{aligned}
\end{equation}
where $ t_{\rm lb}(z) $ is the look-back time for binaries formed at redshift $ z $, $ M_{*} \simeq 0.55\,\msun$ the mean mass of a stellar system in population with the \cite{Kroupa2001MNRAS.322..231K} IMF, $f_{\rm bin}$ = 0.7 the fraction of stars in binaries \citep{Sana2012Sci...337..444S}, $W_{\rm b}$ the contribution of specific binaries we are interested in \citep{Hurley2002MNRAS.329..897H}, $ \mathcal{SFR}(z) $ the cosmic star formation rate density as a function of $ z $, usually peaked at $ z\sim 1.9 $ and declined exponentially at later time. We assume that in our models the star formation commenced at $ z $ = 15.

The BBH progenitor binaries formed at redshift $ z_{\rm f} $ would merge as GW sources at $ \zm $ ($ \zm < z_{\rm f}$) after a delay time $ t_{\rm delay} $, which is defined as the interval between the formation of the progenitor binary and the coalescence of the BBH, i.e., $ t_{\rm delay}=t_{\rm inspiral} +T$, where $ T< 10 \;\rm Myr$ is the lifetime of progenitor system, if the orbital angular momentum loss is efficient enough. So we can get the look-back times at their formation
\begin{equation}
	t_{\rm lb}(z=z_{\rm f}) = \tau_{\rm H} \int_{0}^{z_{\rm f}} \frac{1}{(1+z)E(z)} \rm{d}z,
	\label{eq:tlb}
\end{equation}
where $E(z) = [\Omega_{\rm m}(1+z)^{3}+\Omega_{\lambda}]^{1/2}$, and at their merger
\begin{equation}
	t_{\rm merg} = t_{\rm lb}(z=z_{\rm m}) = t_{\rm lb}(z=z_{\rm f}) -t_{\rm delay}.
	\label{eq:t_merg}
\end{equation}
In our calculation, we employ the flat $ \rm \Lambda CDM $ model with $ H_0=67.8\, \rm kms^{-1}Mpc^{-1} $, $ \Omega_{\rm m}=0.3 $ and $ \Omega_{\rm \lambda} =0.7 $, where $ \tau_{\rm H} = 1/H_0=14.4\, \rm Gyr $ is the Hubble time \citep{Planck2016A&A...594A..13P}.
We adopt the cosmic $ \mathcal{SFR}(z) $ density in \cite{Madau2014ARA&A..52..415M}:
\begin{equation}
	\mathcal{SFR}({z}) = \frac{0.015(1+z)^{2.7}}{1+((1+z )/2.9)^{5.6}}\, \msun \rm yr^{-1}Mpc^{-3},
	\label{eq:sfr}
\end{equation}
and the metallicity as a function of redshift $z$ in \cite{Belczynski2016Natur.534..512B}. For the portions of distributions extending beyond the metallicity range $ [0.0002, 0.02]$, we use the recorded information of the systems at $ Z=0.0001 $ or $0.02$. We exclude the mergers in the near future ($t_{\rm merg}<0$).

\subsection{Character strain} \label{sub:detection}
The characteristic strain of the GW signals at the $n$th harmonic can be calculated following \cite{Kremer2019PhRvD..99f3003K},
\begin{equation}
		h_{c,n}^2 = \frac{2}{3 \pi^{4/3}} \frac{G^{5/3}}{c^3} \frac{M_{\mathrm{c},z}^{5/3}}{D_L^2} \frac{1}{f_{n,z}^{1/3} \left(1+z\right)^2}\left(\frac{2}{n}\right)^{2/3} \frac{g(n,e)}{F(e)},
	\label{eq:hcn}
\end{equation}
where $M_{c,z} = M_c(1+z) = \frac{(m_1m_2)^{3/5}}{(m_1+m_2)^{1/5}}(1+z) $ is the observed chirp mass at redshift $ z $, and $D_L$ is the luminosity distance to the source calculated by
\begin{equation}
	D_L(z)=\frac{c(1+z)}{H_0}\int_0^{z} \frac{dz'}{E(z')},
	\label{eq:dl}
\end{equation}
$f_{n,z}=\frac{f_n}{1+z}=\frac{nf_{\rm orb}}{1+z} $ is the observed frequency of the $n$th harmonic ($f_n$ is the frequency of the $n$th harmonic in the source frame and $f_{\rm orb}$ is the source-fame orbital frequency), $ g(n,e) $ is the function of eccentricity, and $ F(e) $ is the eccentricity correction factor defined to be \citep{Peters1963PhRv..131..435P}:
\begin{equation}
	F(e) = \sum_{n=1}^{\infty} g(n,e)
	=  \frac{1}{\left ( 1 - e^2 \right)^{7/2}} \left ( 1 + \frac{73}{24} e^2 + \frac{37}{96} e^4 \right).
	\label{eq:Fe}
\end{equation}
As the GW power is sharply peaked at the peak frequency $f_{\rm peak}$ \citep{Peters1963PhRv..131..435P}, we calculate the characteristic strain of our modeled GW sources at the peak frequency $f_{\rm peak} $ for simplicity \citep{Hamers2021RNAAS...5..275H},
	\begin{equation}
		\begin{aligned}
		f_{\rm peak} =  & \frac{\sqrt{G \left( m_1+m_2 \right)}}{\pi} \times \\
		&\frac{1 - 1.01678e + 5.57372e^2 - 4.9271e^3 + 1.68506e^4}{ \left[ a \left ( 1-e^2 \right ) \right ]^{1.5}}.
		\end{aligned}
		\label{eq:fpeak}
	\end{equation}
Thus $n$ = $n_{\rm peak} = f_{\rm peak}/f_{\rm orb}$ \citep{Wang2022MNRAS.515.5106W}.

During the inspiral, the eccentricity changes due to gravitational radiation \citep{Peters1964PhRv..136.1224P},
\begin{equation}\label{eq:e(t)}
	{de \over dt}= -{19 \over 12}{\beta \over c_0^4}{e^{-29/19}(1-e^2)^{3/2} \over \left[1+{121 \over 304} e^2
		\right]^{1181/2299}},
\end{equation}
and the orbital separation evolves with eccentricity
\begin{equation}\label{eq:a(e)}
	a(e)= {c_0 e^{12/19} \over (1-e^2)} \left[1+{121 \over 304} e^2
	\right]^{870/2299},
\end{equation}
where $c_0$ is determined by the initial condition $a(e_0) = a_0$ (see Eq. [\ref{eq:c0}]).

\begin{table*}
	\begin{center}
		\caption{Numbers of GW190521-like systems evolved via each evolution channel from ZAMS binaries to BBHs. ``MT+CE": system experiences stable mass transfer (via RLOF or wind mass loss) and later once CE phase, ``MT+MT": system without CE evolution, ``CE+ MT": system experiences once CE phase and later stable RLOF or wind accretion, ``CE+CE": system experiences twice CE phases. The corresponding minimum and maximum of $v_{\rm kick,1}$ and $v_{\rm kick,2}$ are also shown followed their numbers.}
		\renewcommand\arraystretch{1.5}
		\begin{tabular}{cccccc}
			\hline\hline
			$\alpha_{\rm CE}$&$kick_{\rm F}$ & MT+CE &  MT+MT & CE+MT & CE+CE \\
			&$v_{\rm kick}$ [$\rm kms^{-1}$] & &   &  & \\\cline{1-6}
			&$ k1 $&- &19 &- &- \\
			&$ v_{\rm kick,1},v_{\rm kick,2} $& &0, 41.305$-$63.905 & & \\\cline{3-6}
			0.5&$ k2 $&- &32 &- &- \\
			&$ v_{\rm kick,1},v_{\rm kick,2} $& &0, 36.9$-$80.6 & & \\\cline{3-6}
			&$ k3 $&17279  &26 &- &-\\
			&$ v_{\rm kick,1},v_{\rm kick,2} $&1.695$-$46.487, 0.004$-$691.626 & 0.015$-$40.476, 16.843$-$114.77 & &\\\cline{2-6}	
			
			&$ k1 $ &- &19 &- &-\\
			&$ v_{\rm kick,1},v_{\rm kick,2} $ & &0, 41.305$-$63.905& &\\\cline{3-6}			
			1.0 &$ k2 $ &- &32 &- &-\\
			$\ei=0$ &$ v_{\rm kick,1},v_{\rm kick,2} $& &0, 36.9$-$80.6 & &\\\cline{3-6}
			&$ k3 $ &14862 &28 &- &-\\
			&$ v_{\rm kick,1},v_{\rm kick,2} $ &2.499$-$46.499, 0.028$-$691.626 &3.942$-$40.476, 21.428$-$383.477 & &\\\cline{2-6}
			
			&$ k1 $ &- &20 &- &-\\
			&$ v_{\rm kick,1},v_{\rm kick,2} $& &0, 41.305$-$63.905 & & \\\cline{3-6}
			3.0 &$ k2 $ &- &32 &- &-  \\
			&$ v_{\rm kick,1},v_{\rm kick,2} $ & &0, 36.9$-$80.6 & &  \\\cline{3-6}
			&$ k3  $&8671 &24 &-  &- \\
			&$ v_{\rm kick,1},v_{\rm kick,2} $&1.969$-$46.605, 0.111$-$693.735 &2.265$-$44.953, 19.369$-$214.626 &  & \\\hline			
			
			&$ k1 $&- &11 &134 &-\\	
			&$ v_{\rm kick,1},v_{\rm kick,2} $& &0, 36.286$-$72.81 &0, 171.063$-$394.826 &\\\cline{3-6} 			    		
			0.5&$ k2 $&- &22 &105 &-\\
			&$ v_{\rm kick,1},v_{\rm kick,2} $& &0, 30.589$-$90.388 &0, 166.311$-$413.726 &\\\cline{3-6}
			&$ k3 $&18394 &17 &235 &-\\
			&$ v_{\rm kick,1},v_{\rm kick,2} $&0.379$-$46.509, 0.092$-$697.315 &4.734$-$45.848, 34.513$-$142.709  &0.092$-$44.373, 99.196$-$422.095 &-\\\cline{2-6}
			
			&$ k1 $ &- &10 &45 &-\\
			&$ v_{\rm kick,1},v_{\rm kick,2} $& &0, 36.286$-$72.81 &0, 146.125$-$249.277 &\\\cline{3-6}
			1.0 &$ k2 $ &- &24 &63 & -\\
			$\ei=0\sim1$&$ v_{\rm kick,1},v_{\rm kick,2} $& &0, 30.589$-$90.388 &0, 134.284$-$301.761 & \\\cline{3-6}
			&$ k3 $ &15887 &26 &63 &-\\
			&$ v_{\rm kick,1},v_{\rm kick,2} $& 0.149$-$46.187, 0.032$-$687.68 &1.202$-$46.288, 19.738$-$152.097 &0.021$-$41.708, 102.228$-$260.19 &\\\cline{2-6}
			
			&$ k1 $ &- &11 &19 &- \\
			&$ v_{\rm kick,1},v_{\rm kick,2} $& &0, 36.286$-$72.81 &0,104.945$-$171.183 & \\\cline{3-6}
			3.0 &$ k2 $ &-&23 &22 &-\\
			&$ v_{\rm kick,1},v_{\rm kick,2} $& &0, 30.589$-$90.388 & 0, 109.604$-$186.818 & \\\cline{3-6}
			&$ k3  $&9250 &15  &28 &- \\
			
			&$ v_{\rm kick,1},v_{\rm kick,2} $&0.899$-$46.42, 0.017$-$675.649 &10.485$-$45.848, 19.738$-$190.988 &2.076$-$41.044, 59.636$-$247.057 & \\
			\hline\hline
		\end{tabular}
		\label{table4}
	\end{center}
\end{table*}

\begin{table*}
	\begin{center}
		\renewcommand\arraystretch{1.5}
		\caption{The inferred parameters of GW190521-like systems (merged within $z=1.1$) and their progenitors. The values and uncertainties of each parameter indicate the $50th$, $16th$, and $84th$ percentiles of the posterior samples. $\mdi$, $\mci$, $\ai$ and $\ei$ are the initial component masses, orbital semi-major axis and eccentricity of GW190521 like system's progenitors. $m_1$, $m_2$, $a_0$ and $e_0$ are those of the GW190521-like systems'. $v_{\rm kick,1}$ ($v_{\rm kick,2}$) the natal kick velocities of the first (second) formed BHs and $t_{\rm delay}$ the delay time of GW190521-like systems. Runs for $kick_{\rm F}=k3$.}
		\begin{subtable}{0.8\textwidth}
			\begin{tabular}{ccccccc}
				\hline\hline
				Model & $\alpha_{\rm CE}$ & $\mdi~(\msun)$ & $\mci~(\msun)$ & $\log\ai~(\rsun)$ & $\ei$& $\log a_{\rm 0}~(\rsun)$ \\\hline
				& 0.5 &$433.71_{-55.03}^{+147.60}$ &$42.95_{-2.42}^{+3.01}$ &$4.16_{-0.28}^{+0.39}$&0 &$1.89_{-0.08}^{+0.17}$ \\
				$\ei=0$&1.0&$419.53_{-46.43}^{+198.54}$& $42.80_{-3.14}^{+3.68}$ &$4.19_{-0.28}^{+0.36}$&0 &$1.88_{-0.07}^{+0.10}$ \\
				&3.0&$444.77_{-67.40}^{+224.67}$ &$37.83_{-2.79}^{+4.08}$ &$4.16_{-0.24}^{+0.35}$ &0& $1.84_{-0.07}^{+0.09}$  \\\hline
				&0.5&$450.16_{-69.61}^{+232.91}$ & $43.07_{-3.13}^{+3.11}$ &$4.19_{-0.29}^{+0.34}$&$0.49_{-0.32}^{+0.24}$ &$1.90_{-0.09}^{+0.54}$\\
				$\ei=0\sim 1$&1.0&$422.61_{-47.97}^{+206.80}$ &$43.18_{-3.38}^{+3.33}$ &$4.22_{-0.28}^{+0.37}$ &$0.47_{-0.33}^{+0.24}$&$1.88_{-0.07}^{+0.15}$\\
				&3.0&$442.68_{-67.86}^{+229.05}$ &$37.71_{-2.57}^{+4.36}$ &$4.22_{-0.29}^{+0.44}$ &$0.46_{-0.28}^{+0.27}$& $1.84_{-0.07}^{+0.11}$\\
				\hline\hline
			\end{tabular}
		\end{subtable}
		
		\begin{subtable}{0.8\textwidth}
			\begin{tabular}{cccccc}
				
				\hline\hline
				$e_{0}$ &$m_1({\msun})$& $m_2({\msun}$) &$v_{\rm kick,1}~(\rm kms^{-1})$ &$v_{\rm kick,2}~(\rm kms^{-1})$ & $t_{\rm delay}~(\rm Gyr)$ \\\hline
				$0.35_{-0.25}^{+0.29}$ & $164.14_{-10.22}^{+8.36}$ & $17.26_{-1.52}^{+1.58}$ & $33.77_{-11.27}^{+5.90}$ &$229.66_{-165.78}^{+109.71}$ & $5.65_{-1.84}^{+3.24}$\\
				$0.27_{-0.18}^{+0.31}$&$163.55_{-8.30}^{+9.28}$&$17.15_{-1.82}^{+2.00}$&$34.03_{-10.45}^{+6.13}$&$180.34_{-117.76}^{+165.35}$&$5.94_{-2.09}^{+3.09}$\\
				$ 0.24_{-0.17}^{+0.27} $&$ 164.13_{-9.38}^{+9.84} $&$ 14.61_{-2.67}^{+1.93} $&$ 33.83_{-8.96}^{+6.22} $&$ 186.98_{-134.74}^{+163.48} $&$ 5.81_{-2.06}^{+3.14} $\\\hline
				$ 0.38_{-0.27}^{+0.51} $&$ 165.51_{-11.03}^{+8.21} $&$ 17.24_{-1.56}^{+1.61} $&$ 27.56_{-12.94}^{+10.36} $&$ 220.53_{-143.98}^{+119.35} $&$ 5.76_{-2.00}^{+2.94} $\\
				$ 0.28_{-0.19}^{+0.38} $&$ 165.05_{-8.83}^{+8.21} $&$ 17.32_{-1.82}^{+1.61} $&$ 28.94_{-13.62}^{+8.70} $&$ 184.35_{-121.13}^{+160.26} $&$ 5.92_{-2.05}^{+3.08} $\\
				$ 0.24_{-0.17}^{+0.32} $&$ 164.34_{-9.21}^{+10.80} $&$ 14.62_{-2.45}^{+2.10} $&$ 27.68_{-14.17}^{+10.00} $&$ 173.27_{-122.20}^{+201.65} $&$ 5.80_{-2.05}^{+3.13} $\\
				\hline\hline
			\end{tabular}
		\end{subtable}
		
		\label{table5}
	\end{center}
\end{table*}	
	
\begin{table*}
	\begin{center}
		\caption{The calculated merger rate density of GW190521-like systems with the natal kick drawn from a Maxwell distribution of NS kick velocity ($\sigma_{\rm NS}=265$ kms$^{-1}$). The 3rd-6th columns present the numbers of BBHs with the progenitor systems formed at relevant metallicity and merged in local universe with $\zm\leq 0.48$ and 1.1 (in parentheses) per $\rm Gpc^3$ per $\rm year$. The last column lists the cumulated  merger rate densities taking into account cosmic evolution.}
		\renewcommand\arraystretch{1.5}
		\begin{tabular}{cccccccc}
			\hline\hline
			$\alpha_{\rm CE}$&$kick_{\rm F}$ &$Z=0.0002$&  $Z=0.0004$& $Z=0.0008$& $Z=0.0016$ &$\mathcal{R}(\zm\leq 0.48)$($(\zm \leq 1.1)$)\\
			 && & & &&[$\rm Gpc^{-3}yr^{-1}$] \\\hline
			 &&\multicolumn{5}{c}{$\ei=0$}&\\\cline{3-7}
			 &$ k1 $&-\;(-) &-\;(-) &-\;(-) &4.067e-05\;(4.067e-05)&4.067e-05\;(4.067e-05)\\\cline{3-7}
			0.5&$ k2 $&9.087e-05\;(2.111e-04) &1.180e-04\;(3.193e-04) &3.089e-04\;(6.838e-04)  &7.771e-05\;(3.457e-04)&5.955e-04\;(1.560e-03)\\\cline{3-7}
			 &$ k3 $&2.668e-04\; (1.738e-03) &2.498e-03\;(8.888e-03)  &3.847e-03\; (1.583e-02) &1.192e-03\; (5.759e-03)&7.803e-03\;(3.221e-02)\\\cline{2-7}
			
		   &$ k1 $ &-\;(-) &-\;(-) &-\;(-) &6.889e-05\;(6.889e-05)&6.889e-05\;(6.889e-05)\\\cline{3-7}
			1.0 &$ k2 $ &9.087e-05\;(2.112e-04) &1.180e-04\;(3.193e-04) &3.089e-04\;(6.838e-04) &7.771e-05\;(3.457e-04)&5.955e-04\;(1.560e-03)\\\cline{3-7}
		  &$ k3 $ &1.380e-03\; (5.235e-03) &2.498e-03\; (9.151e-03) &4.352e-03\; (1.516e-02) &2.099e-03\; (7.390e-03)& 1.033e-02\;(3.694e-02)\\\cline{2-7}

			 &$ k1 $ &-\;(-) &-\;(-) &-\;(-) &5.724e-05\;(5.724e-05)&5.724e-05\;(5.724e-05)\\\cline{3-7}
			3.0 &$ k2 $ &9.087e-05\;(2.112e-04) &1.180e-04\;(3.193e-04) &3.089e-04\;(6.838e-04) &7.771e-05\;(3.457e-04)&5.955e-04\;(1.560e-03)\\\cline{3-7}
			 &$ k3 $&1.352e-03\; (4.831e-03) &2.183e-03\; (8.117e-03) &4.267e-03\; (1.602e-02) &6.142e-03\; (2.230e-02)&1.394e-02\;(5.127e-02)\\\hline
		    &&\multicolumn{5}{c}{$\ei=0\sim1$}&\\\cline{3-7}
			 &$ k1 $ &-\;(-) &-\;(-) &-\;(-) &2.841e-03\;(8.544e-03)&2.841e-03\;(8.544e-03)\\\cline{3-7}
			0.5 &$ k2 $ &1.639e-05\;(9.837e-05) &0.0\;(1.582e-04) &5.693e-05\;(5.693e-05) &1.0750e-03\;(3.023e-03)&1.148e-03\;(3.337e-03)\\\cline{3-7}
			 &$ k3 $ &2.431e-04\; (2.727e-03) &2.392e-03\; (9.022e-03) &4.323e-03\; (1.637e-02) &4.364e-03\; (1.726e-02) &1.132e-02\;(4.537e-02)\\\cline{2-7}
			 &$ k1 $ &-\;(-) &-\;(-) &-\;(-) &1.208e-03\;(4.552e-03)&1.208e-03\;(4.552e-03)\\\cline{3-7}
			1.0&$ k2 $ &1.639e-05\;(9.837e-05) &0.0\;(1.582e-04) &5.693e-05\;(5.693e-05)& 4.771e-04\;(1.449e-03)& 5.550e-04\;(1.763e-03)\\\cline{3-7}
			 &$ k3 $ &1.422e-03\; (5.271e-03) &2.354e-03\; (8.712e-03)  &4.438e-03\; (1.632e-02) &2.927e-03\; (1.445e-02)& 1.114e-02\;(4.476e-02)\\\cline{2-7}
			 &$ k1 $ &-(-) &-(-) &-\;(-)&3.320e-04(1.071e-03) &3.320e-04\;(1.071e-03)\\ \cline{3-7}					
			3.0 &$ k2 $&1.639e-05\;(9.837e-05)  &0.0\;(1.582e-04) &5.693e-05\;(5.693e-05) &1.048e-03\;(2.369e-03)&1.121e-03\;(2.683e-03)\\\cline{3-7}
			 & $ k3 $&1.370e-03\; (4.888e-03) &2.141e-03\; (7.988e-03)  &4.024e-03\; (1.549e-02) &6.149e-03\; (2.249e-02) &1.369e-02\;(5.085e-02)\\
			\hline\hline
		\end{tabular}
		\label{table6}
	\end{center}
\end{table*}

\section{Results} \label{sec:results}
\begin{figure}
	\centering
	\includegraphics[width=1.0\linewidth]{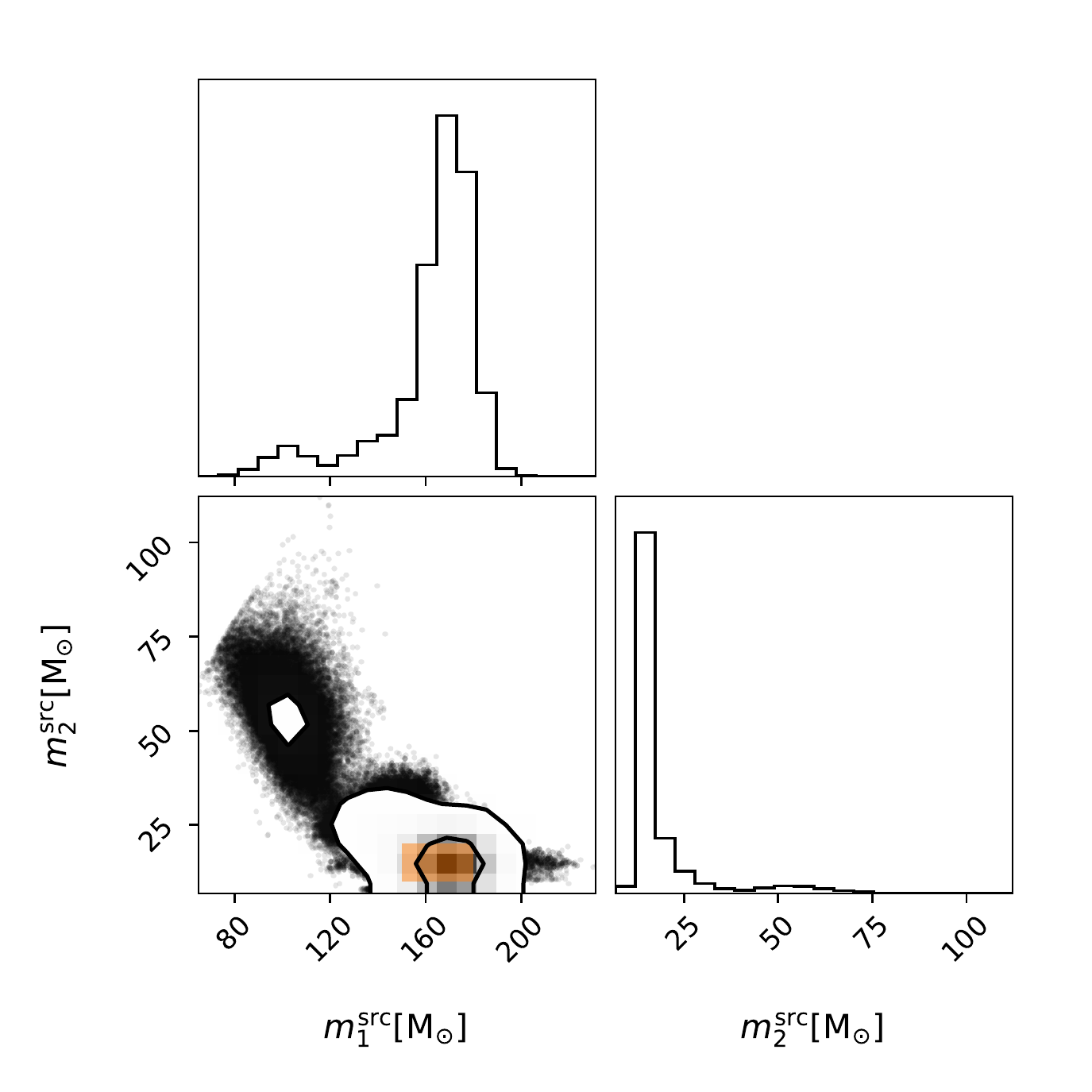}
	\caption{The posterior distribution of GW190521 from \protect\cite{Nitz2021ApJ...907L...9N} under $Prior_{q-M}$ prior, overlaid yellow region is the component masses of our calculated GW190521-like systems.}
	\label{nitz}
\end{figure}

For each model we regard the merging BBHs with component masses $m_1=[150:180]\,\msun$ and $m_2=[10:20]\,\msun$ as GW190521-like systems (as shown in Fig.\ref{nitz}). According to the analysis of \cite{Nitz2021ApJ...907L...9N}, the primary BH mass in GW190521 is $\sim 170\,\msun$. We simulate binary evolution with the primary mass $\leq 900\,\msun$ at different metallicities and find that such massive BHs can only form at relatively low metallicities ($Z\leq 0.0016$), with the pre-collapse core-helium masses heavier than $135\,\msun$, while stars at higher metallicity ($Z\geq 0.0032$) will undergo PISNe, triggering complete disruption of the star and leaving no compact remnants. This is because stars formed in lower metallicity environments can reach higher central temperatures, which results in larger core masses than their counterparts at higher metallicity. So our following discussions only refer to the data set of the models with $Z\leq 0.0016$.

Table \ref{table4} lists the predicted numbers of GW190521-like systems and kick velocity distributions under different conditions. Their main features can be summarized as follows.
\begin{itemize}
	\item Most of GW190521-like systems form through the ``MT+CE" channel. Here, ``MT" means that stars in a binary interact via stable RLOF or wind accretion, and the ``CE" phase is triggered by eccentric collision of both stars at periastron instead of dynamically unstable mass transfer caused by the expansion of the donor star. Binary stars in eccentric orbits may collide at periastron before either one fills its RL, and such collisions lead to CE evolution if at least one of the stars is a giant-like star \citep{Hurley2002MNRAS.329..897H}. Thus only the $k3$ kick prescription that produces non-zero BH1 natal kick velocities works in this channel. Besides, the number of BBH mergers decreases as $\alpha_{\rm CE}$ increases, because larger $\alpha_{\rm CE}$ leads to wider orbits after the CE phase, making it more difficult for the BBHs to merge. To reproduce GW190521-like systems in this channel requrires $v_{\rm kick,1} \simeq 0-50 \,\rm kms^{-1}$ and $v_{\rm kick,2} \simeq 0-700 \,\rm kms^{-1}$.
	\item Only about 0.1\% of GW190521-like systems form through ``MT+ MT" channel. The systems with moderate low eccentricities can avoid collision at periastron until the BBH formation. So this channel is independent of the value of $\alpha_{\rm CE}$. The predicted natal kick velocities are $v_{\rm kick,1}=0$ and $v_{\rm kick,2} \simeq 30-90 \,\rm kms^{-1}$ ($kick_{\rm F}=k1,k2$), and $v_{\rm kick,1} \simeq 0-50 \,\rm kms^{-1}$ and $v_{\rm kick,2} \simeq 16-400 \,\rm kms^{-1}$ ($kick_{\rm F}=k3$).
	\item For very low mass ratio binaries ($q<0.1$), the secondary star usually does not have enough energy to drive off the CE if it is triggered by dynamically unstable MT, so there is no system formed via the ``CE+MT" channel in the `$\ei=0$' model. In the `$\ei=0\sim 1$' model, the non-zero eccentricity makes CE evolution possible just like in the ``MT+CE" channel. The number of surviving systems also decreases with increasing $\alpha_{\rm CE}$. The predicted natal kick velocities are $v_{\rm kick,1}=0$ and $v_{\rm kick,2}\simeq 100-400 \,\rm kms^{-1}$ ($kick_{\rm F}=k1,k2$), and $v_{\rm kick,1}\simeq 0-45 \,\rm kms^{-1}$ and $v_{\rm kick,2} \simeq 60-420 \,\rm kms^{-1}$ ($kick_{\rm F}=k3$).
	\item No GW190521-like system form though the ``CE+CE" channel.
\end{itemize}   

Table \ref{table5} presents the inferred parameters of GW190521-like systems that will merge within $z=1.1$ and their progenitors. As most of them are formed with $kick_{\rm F}=k3$, we only show the results with the $k3$ prescription. The number and its subscripts and superscripts represent the $50th$, $16th$ and $84th$ percentiles of each parameter.  It is seen that the results do not show significant differences in in the `$\ei=0$' and `$\ei=0\sim1$' models.

The analysis of \cite{Abbott2020PhRvL.125j1102A, Abbott2020ApJ...900L..13A} suggested that GW190521 merged at the redshift of $0.82_{-0.34}^{+0.28}$, while \cite{Nitz2021ApJ...907L...9N} predicted  a luminosity distance of $1.06_{-0.28}^{+1.4}\rm Gpc$ ($z\simeq0.21_{-0.05}^{+0.23}$). According to their restrictions on the redshift, we display the calculated merger rate density $\mathcal{R}$ of GW190521-like systems at $z\leq0.48$ and $z\leq1.1$ in Table \ref{table6}, which lie in the range of  $4\times 10^{-5}-5\times 10^{-2} \,\rm Gpc^{-3}yr^{-1}$. As mentioned above, the merger rate density with the $k3$ kick prescription is much more than with the  $k1$ or $k2$ prescription.

\begin{figure*}
	\centering
	\includegraphics[width=1.0\linewidth]{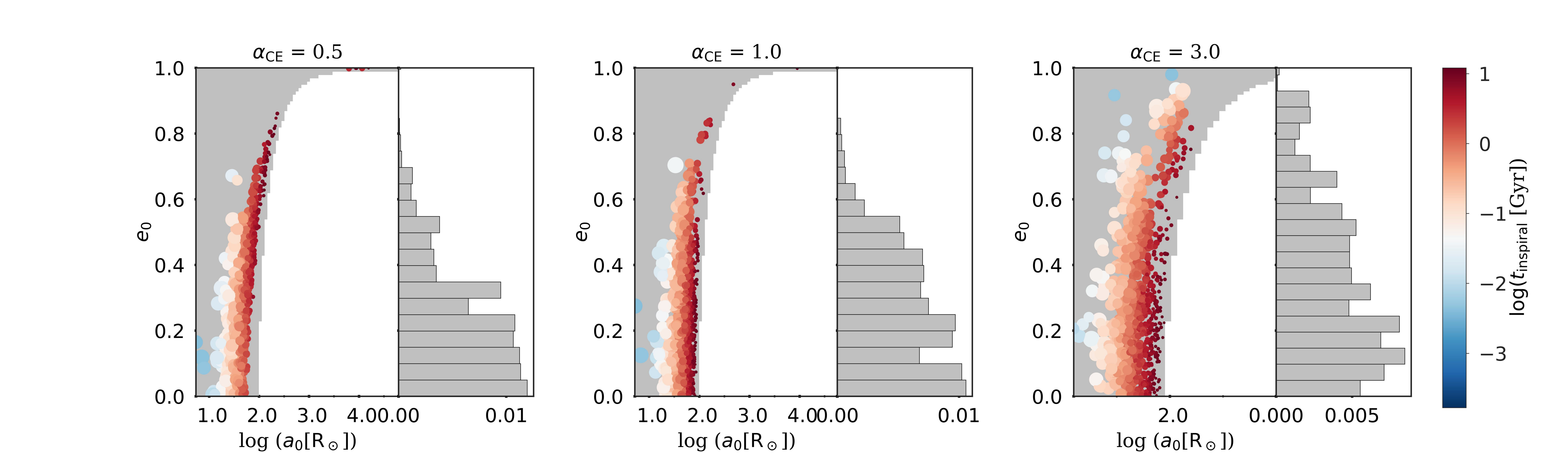}
	\includegraphics[width=1.0\linewidth]{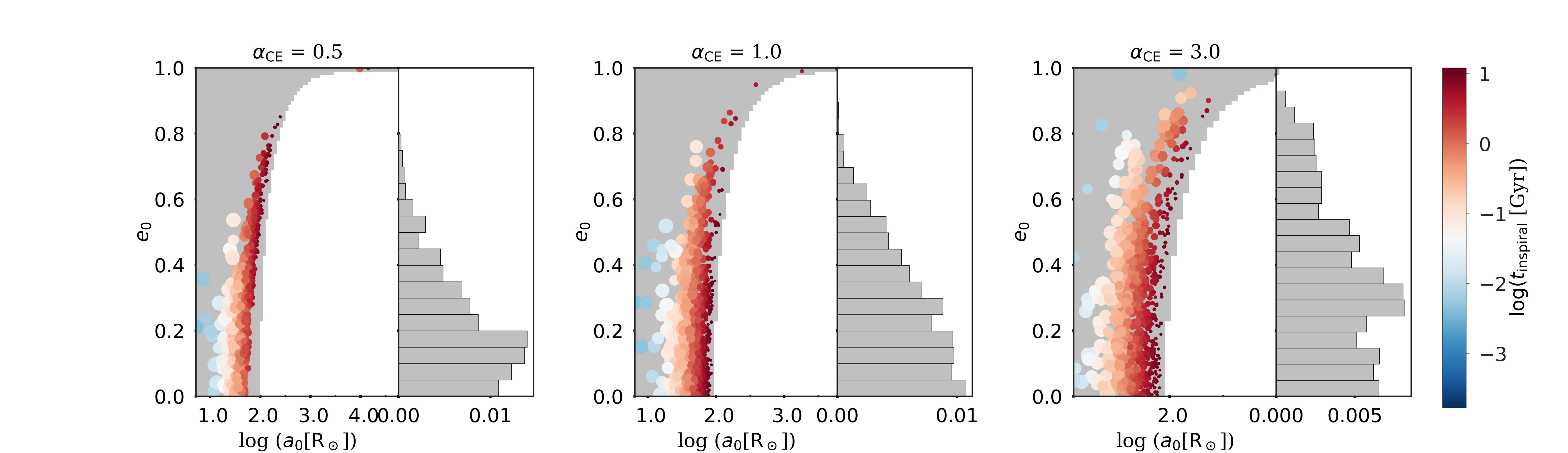}
	\caption{The distribution of the eccentricity versus the orbital semi-major axis at the BBH formation, runs for $kick_{\rm F} = k3$ and metallicity $Z =0.0002$ is shown. The shaded region in each panel represents the area of theoretical parameter space ($e_0\;, a_0$) which satisfies $t_{\rm inspiral}(e_0\;, a_0)\leq\tau_{\rm H}$, while the color-coded scatters label the modeled GW190521-like systems with the colors denoting their $t_{\rm inspiral}(e_0\;, a_0)$ values and the size denotes the weight of each system in the population. Columns from left to right correspond to the simulations with $\alpha_{\rm CE}$ = 0.5, 1.0 and 3.0, respectively. The top and bottom panels correspond to `$\ei=0$' and `$\ei=0\sim1$' models. The horizontal histograms represent the merger rate density $\mathcal{R}$-weighted distribution of $e_0$.}
	\label{z0.0002}
\end{figure*}

Fig.~\ref{z0.0002} shows the distribution of $e_0$ and $a_0$ at the birth of the BBHs (with $Z=0.0002$ and $kick_{\rm F} =k3$). The upper and lower panels correspond to the `$\ei=0$' and `$\ei=0\sim1$' models, and the left, middle, and right panels correspond to $\alpha_{\rm CE}=0.5$, 1.0, and 3.0, respectively. The distribution of $e_0$ tends to be wider with increasing $\alpha_{\rm CE}$ in both `$\ei=0$' and `$\ei=0\sim1$' models. For $\alpha_{\rm CE}=3.0$, there are many BBHs with $e_0>0.8$, while for $\alpha_{\rm CE}$ = 0.5 and 1.0, few BBHs form with extremely eccentric and wide orbits ( $a_0\sim 10^4\rsun$). Most of the BBHs have  $e_0\leq 0.4$ and $a_0$ concentrated within $\sim$ 10 $-$ 100 $\rsun$. BBHs with shorter orbits and moderate eccentricities usually merge earlier than others, as shown by the colorbars.

\begin{figure*}
	\centering
	\includegraphics[width=1.0\linewidth]{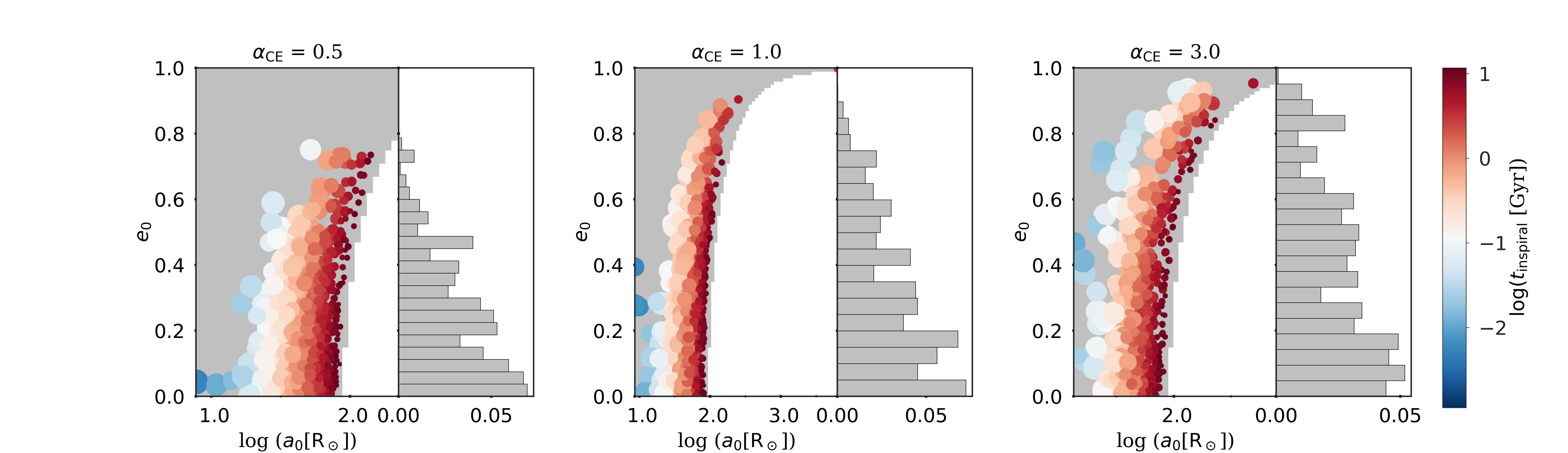}
	\includegraphics[width=1.0\linewidth]{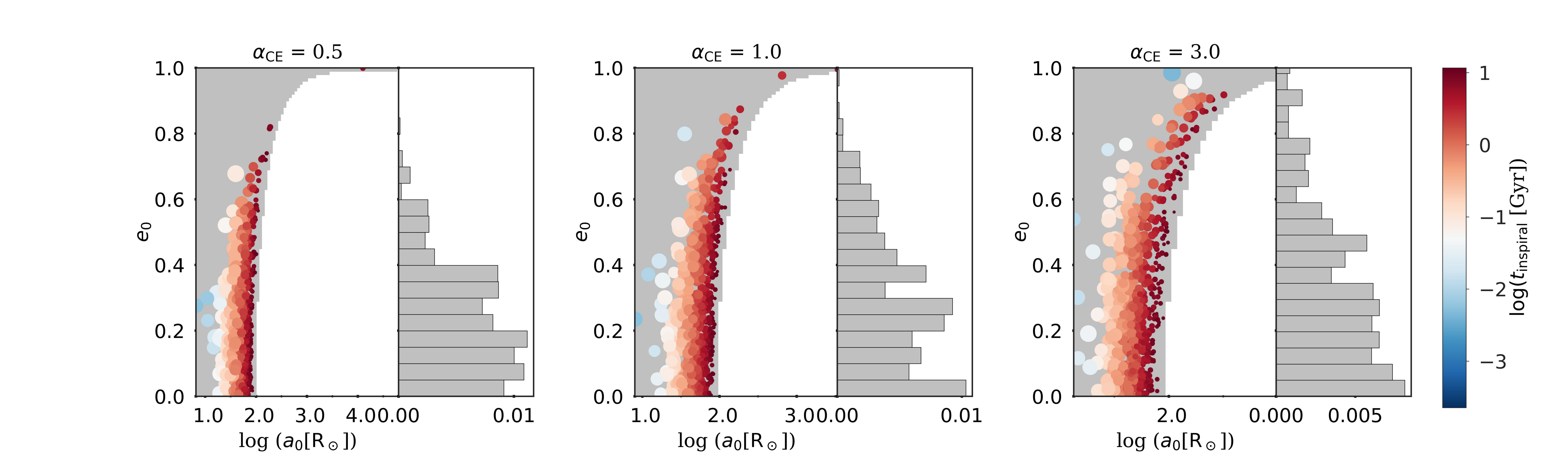}
	\caption{Same as Fig.~\ref{z0.0002}, but for $Z = 0.0004$.}
	\label{z0.0004}
\end{figure*}

\begin{figure*}
	\centering
	\includegraphics[width=1.0\linewidth]{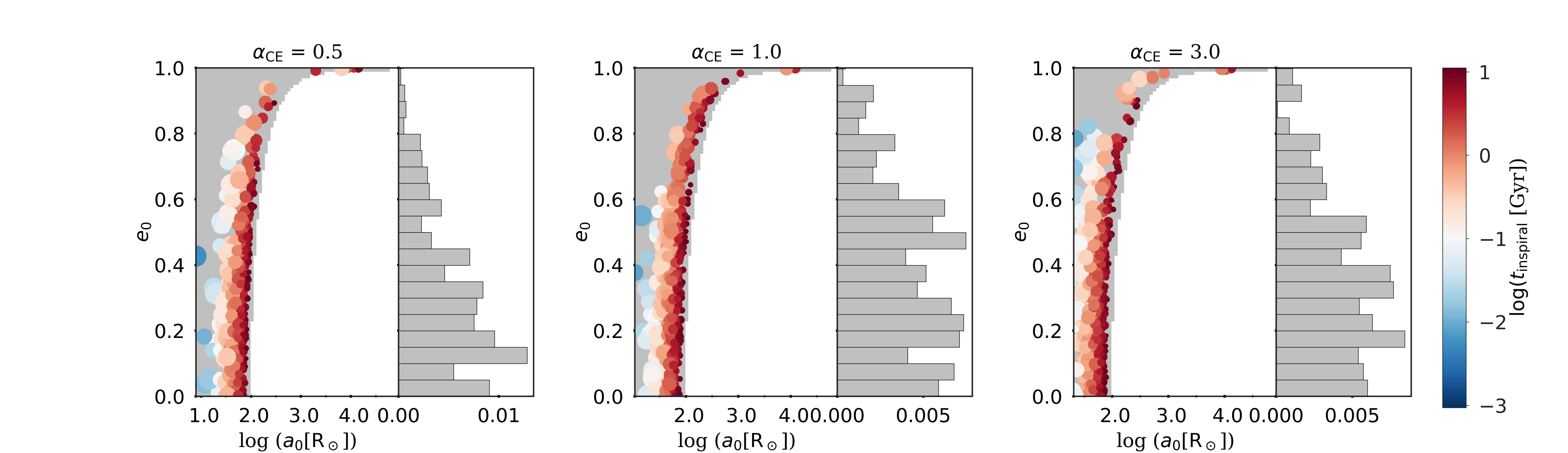}
	\includegraphics[width=1.0\linewidth]{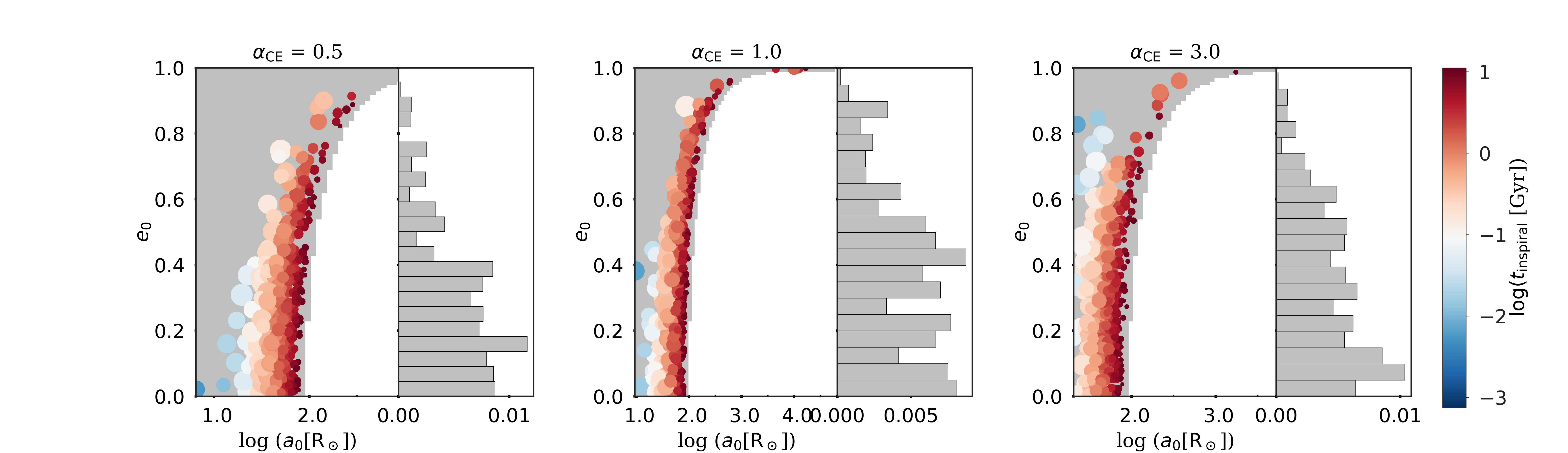}
	\caption{Same as Fig.~\ref{z0.0002}, but for $Z = 0.0008$.}
	\label{z0.0008}
\end{figure*}

\begin{figure*}
	\centering
	\includegraphics[width=1.0\linewidth]{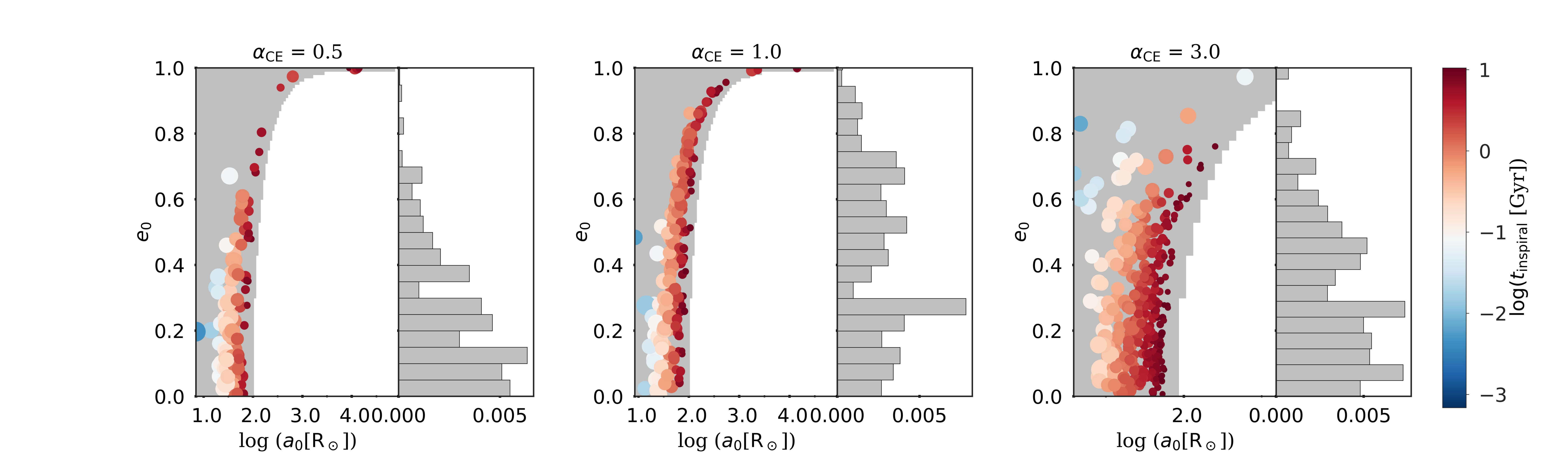}
	\includegraphics[width=1.0\linewidth]{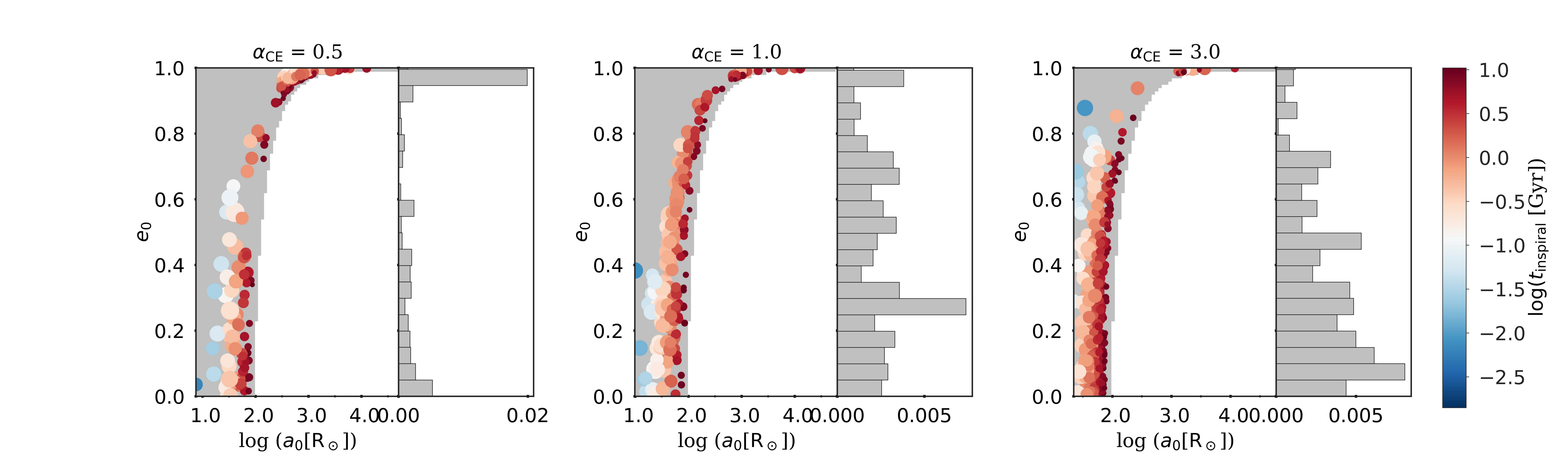}
	\caption{Same as Fig.~\ref{z0.0002}, but for $Z = 0.0016$.}
	\label{z0.0016}
\end{figure*}

Fig.~\ref{z0.0004} $-$ \ref{z0.0016} show the results of $Z=0.0004$, 0.0008 and 0.0016, respectively. They reflect the similar tendency as in Fig.~\ref{z0.0002}. A comparison of Fig.~\ref{z0.0002} $-$ \ref{z0.0016} shows that, as  the metallicity increases, there are more BBHs with large eccentric and wide orbits ($e_0>0.8$ and $a_0>100\rsun$). This is because the progenitors of GW190521-like system at higher $Z$ are more massive and thus have larger size than those at low $Z$.

\begin{figure*}
	\centering
	\includegraphics[width=1.\linewidth]{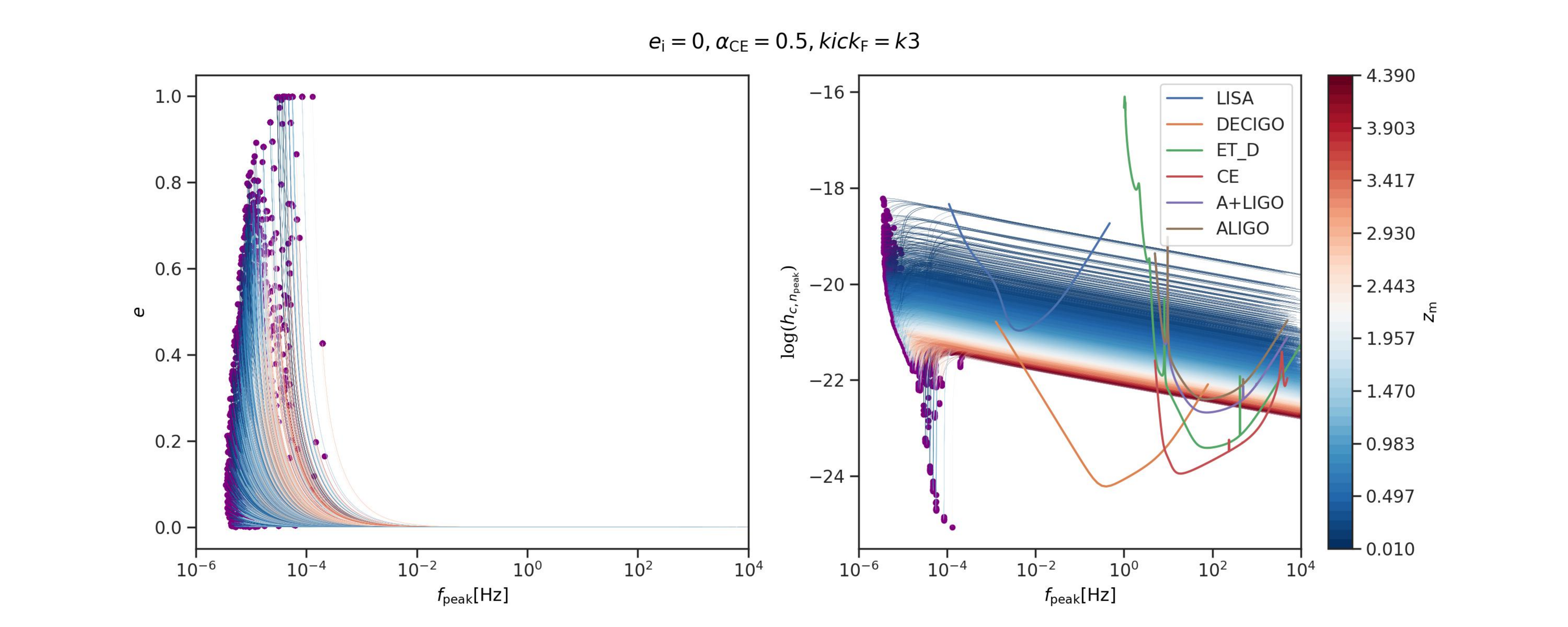}
	\includegraphics[width=1.\linewidth]{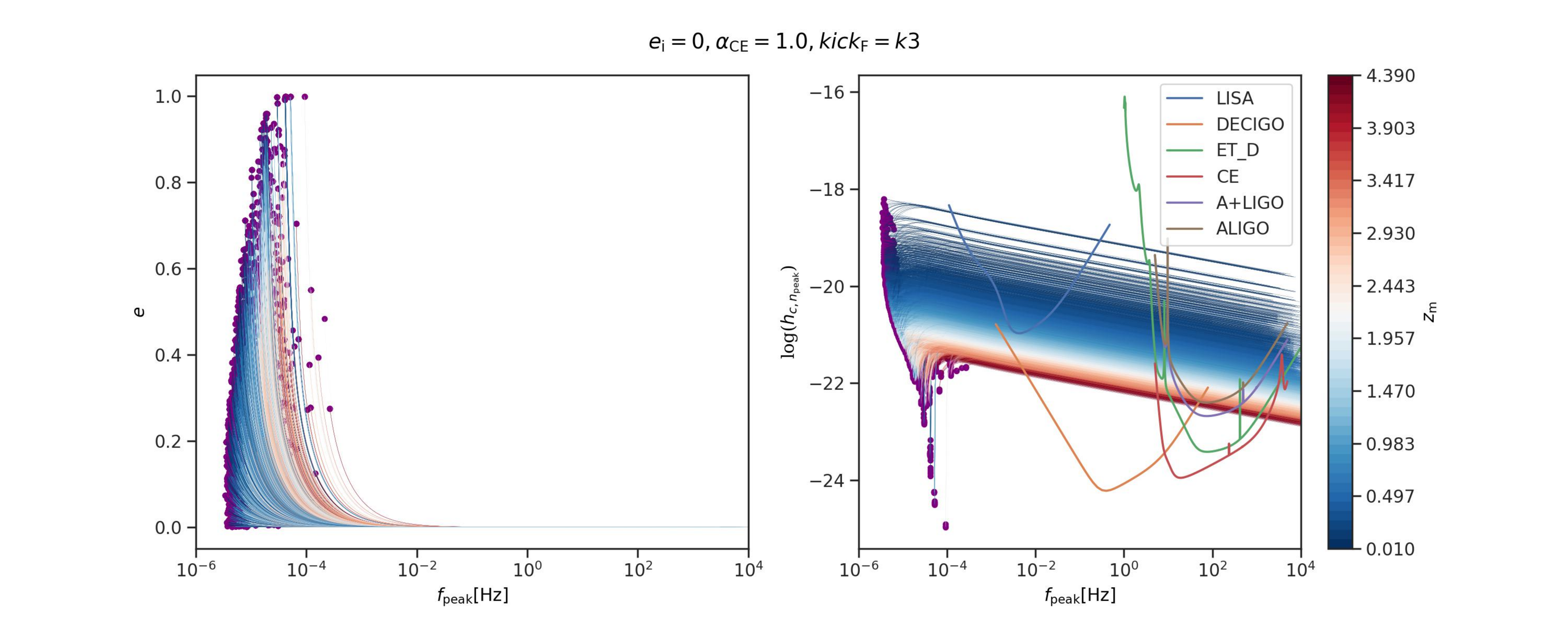}
	\includegraphics[width=1.\linewidth]{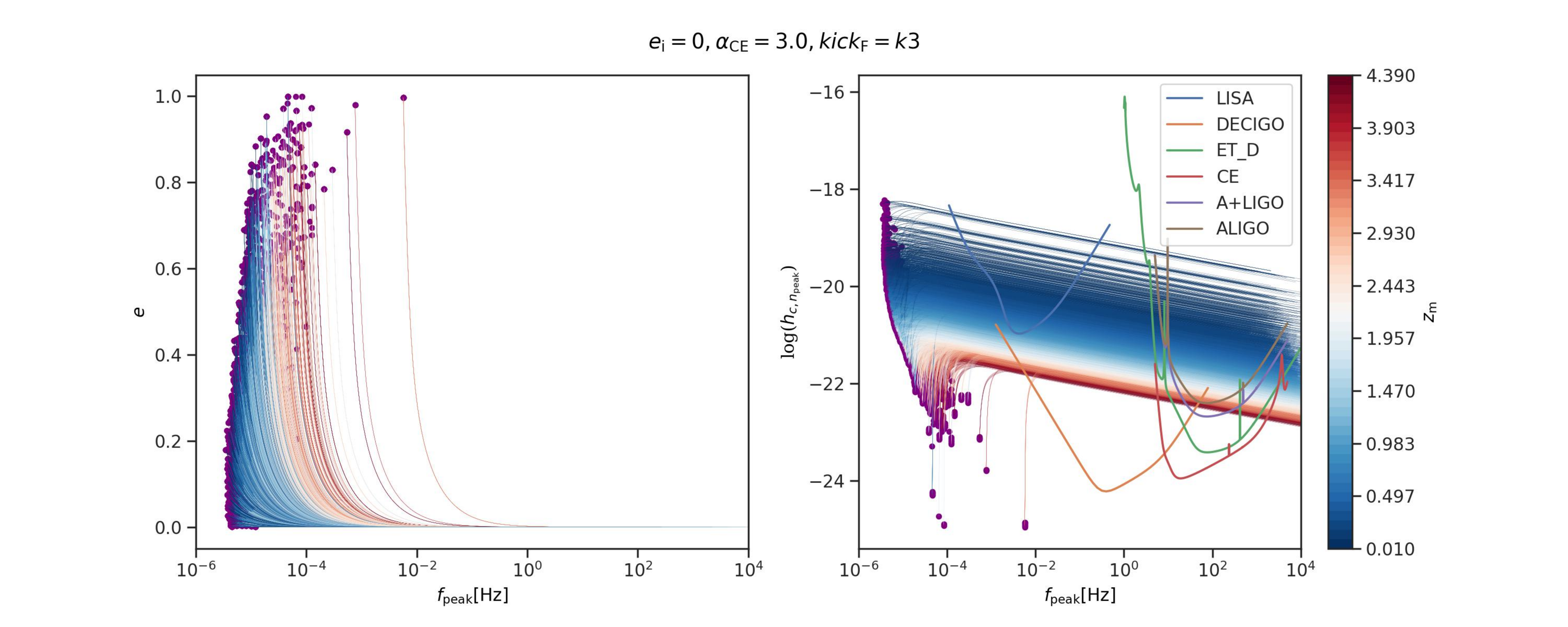}
	\caption{Evolution of the eccentricity (left panels) and the characteristic strain $h_{\rm c,n_{peak}}$ (right panels) versus the peak frequency $f_{\rm peak}$ of GW190521-like systems from inspiral to nearly merger (`$\ei=0$' model, $kick_{\rm F} = k3$). The purple points label the formation of BBHs. From top to bottom are the runs with $\alpha_{\rm CE} = 0.5$, 1.0 and 3.0, respectively. The colors of the line denote the redshift $\zm$ where the merged sources are located. The simulated tracks of the sensitivity curve of space-borne GW detectors LISA \citep{Robson2019CQGra..36j5011R}, DECIGO \citep{Yagi2011PhRvD..83d4011Y} and the ground-based detectors ET \citep{Hild2011CQGra..28i4013H}, CE \citep{Abbott2017CQGra..34d4001A}, A+LIGO and Advanced LIGO\citep{LIGO1,LIGO2} are also shown.}
	\label{chcn}
\end{figure*}

\begin{figure*}
	\includegraphics[width=1.\linewidth]{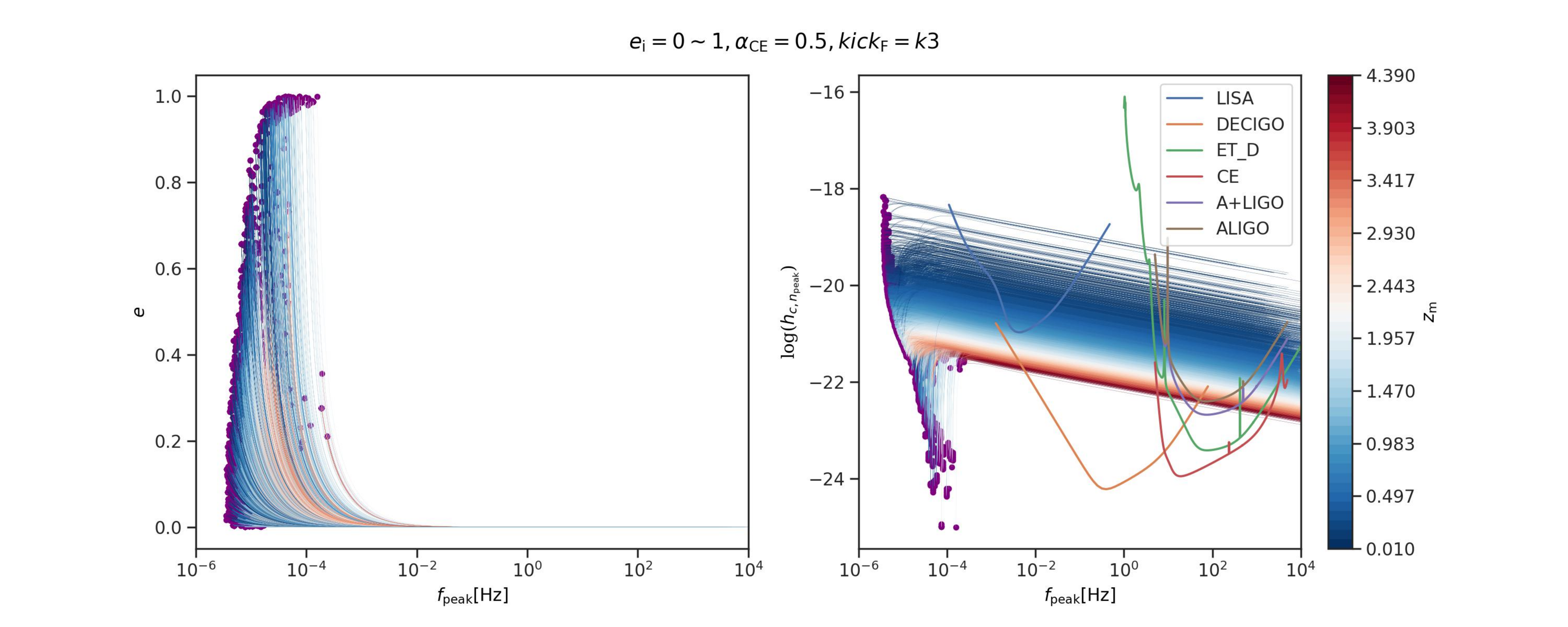}
	\includegraphics[width=1.\linewidth]{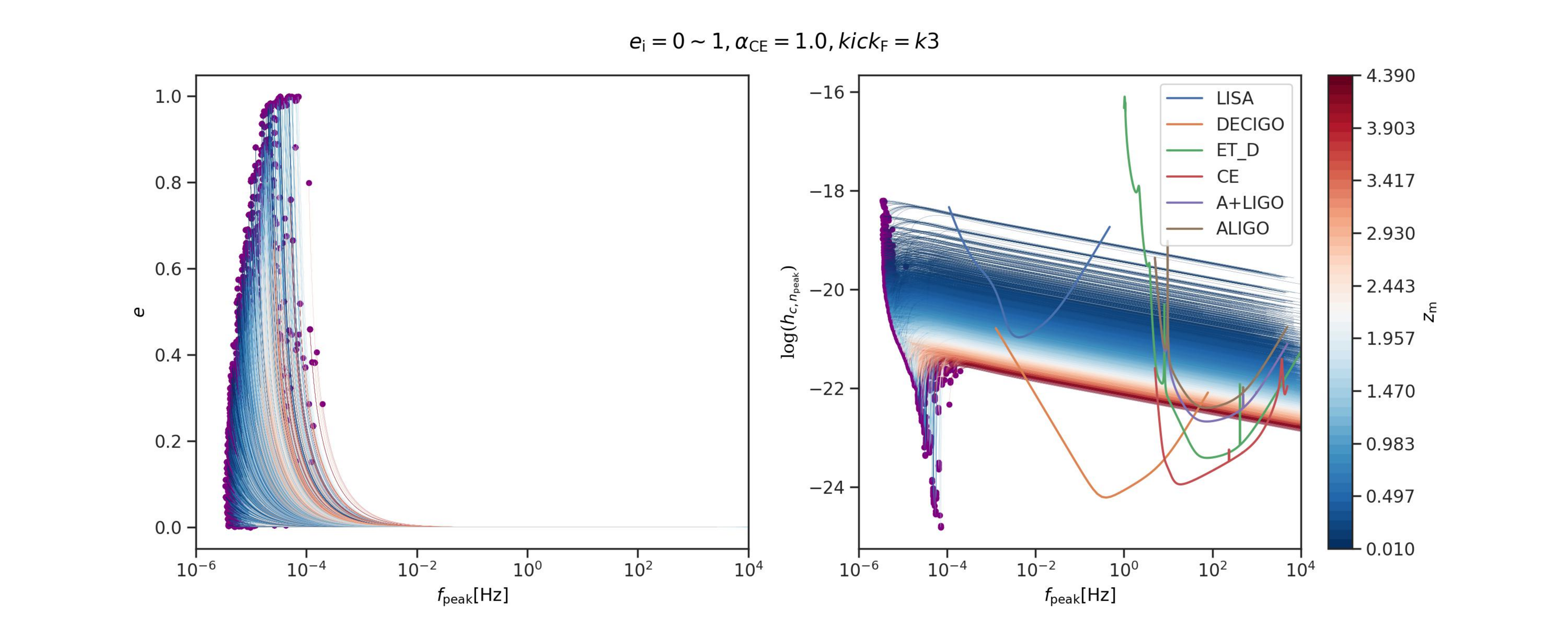}
	\includegraphics[width=1.\linewidth]{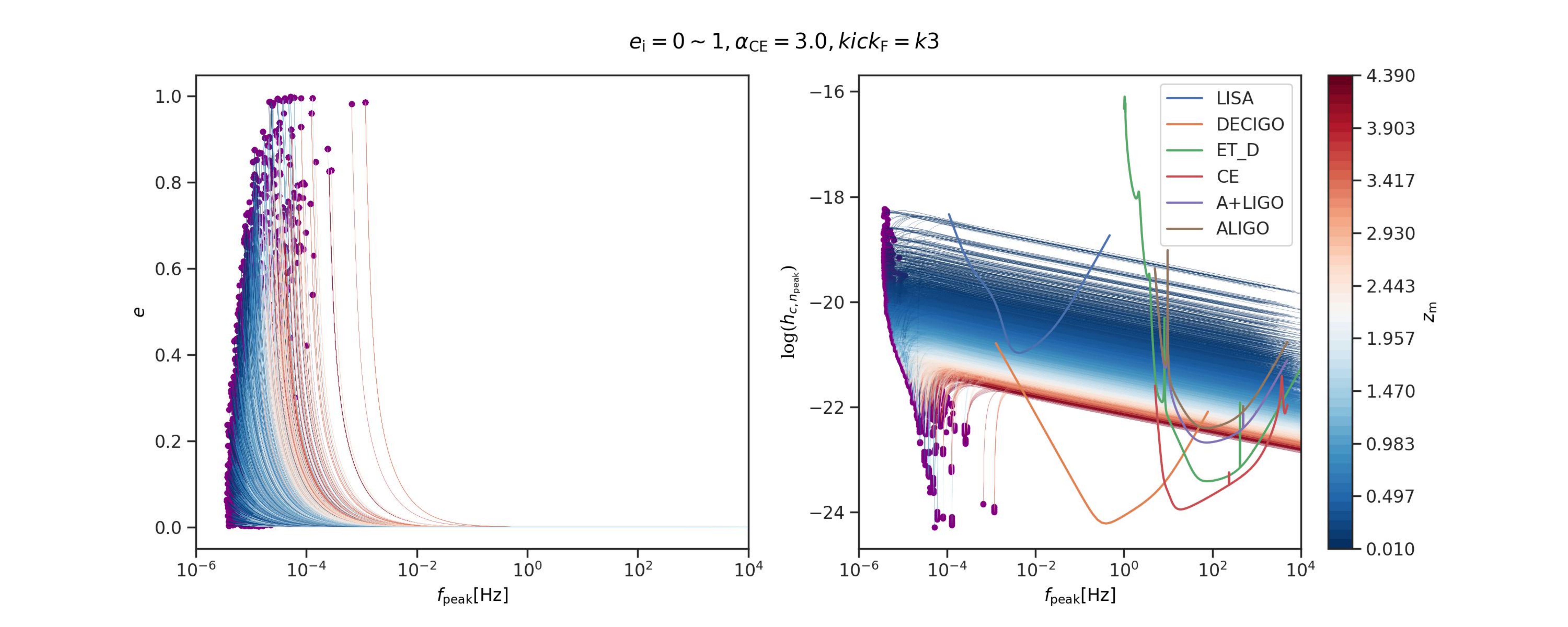}
	\caption{Same as Fig.~\ref{chcn} but for `$\ei=0\sim 1$' model.}
	\label{ehcn}
\end{figure*}

Current and upcoming missions such as the ground-based aLIGO, Cosmic Explorer (CE) \citep{Reitze2019BAAS...51g..35R}, Einstein telescope (ET) \citep{Punturo2010CQGra..27s4002P} and space-borne DECIGO \citep{Seto2001PhRvL..87v1103S} and LISA \citep{Amaro-Seoane2017arXiv170200786A} would detect thousand of merger events of BBHs per year \citep{Evans2021arXiv210909882E}. We explore whether GW190521-like systems could be detected by these instruments. In Fig.~\ref{chcn}, the left three panels present the evolution of the eccentricity of GW190521-like systems during the inspiral prior to merge as a function of the peak frequency. The GW emission dominates the evolution of the binary semi-major axis and eccentricity, leading to efficient circularization of the BBH systems  before the merger. The right three panels present the GW signal characteristic strain at the $n_{\rm peak}th$ harmonic along with the orbital evolution as a function of the peak frequency, which is the key ingredient determining whether such mergers can be seen with the GW detectors, in the `$\ei=0$' model. The evolutionary tracks (calculated with Eqs.~[\ref{eq:e(t)}] and [\ref{eq:a(e)}]) are gradually overlapped by the sensitivity curves of LISA, DECIGO, ET, CE, A+LIGO, and aLIGO during the orbital shrinking and circularizing stage. The systems become largely circularized before entering the sensitivity band of ET, CE, A+LIGO, and aLIGO, and any residual eccentricity is expected to have a negligible effect on their detectability \citep{Mandel2008ApJ...681.1431M}. The most distant detectable GW190521-like mergers are at the redshfit $\sim 4.4$. Fig.~\ref{ehcn} shows the results in the `$\ei = 0\sim1$' model, and there are no significant differences compared with Fig.~\ref{chcn}.

Another important characteristic of merging BBHs is their spins, which get imprinted in the GW signal \citep{Cutler1994PhRvD..49.2658C}. The BH progenitors gain and lose their spin angular momenta through stellar evolution, mass transfer and tidal interactions \citep{Hurley2002MNRAS.329..897H, Belczynski2020A&A...636A.104B, Tanikawa2021ApJ...910...30T}. The spin angualr momenta are generally parallel to the orbital angular momentum, until the kicks to the BHs cause  their spin axes tilted.
According to traditional tidal theroy \citep{Zahn1977A&A....57..383Z, Hut1981A&A....99..126H}, the torque depends on the ratio of the stellar radius $R$ to the separation $a$ of both stars, that is, $\propto (R/a)^6$.
Because the progenitor of BH1 is  very massive and the initial orbit is very wide, spinning up of the BH1's progenitor is ineffective, so the merging BBHs generally have small $|\chi_{\rm eff}|$. In the `$\ei=0$' model, $\chi_{\rm eff} = 0 \sim 0.1$ ($kick_{\rm F}=k3$) and $-0.09\sim 0.1 $ ($kick_{\rm F}=k1, k2$); in the `$\ei=0\sim1$' model, $\chi_{\rm eff} = -0.08 \sim 0.1$ ($kick_{\rm F}=k3$) and $-0.1\sim 0.1$ ($kick_{\rm F}=k1, k2$). They are in contradiction with \cite{Nitz2021ApJ...907L...9N}'s prediction that the BH spin is anti-aligned with the orbital angular momentum and $\chi_{\rm eff}=-0.51_{-0.11}^{+0.24}$.
However, the mechanism of tidal interactions are not well understood. If we adopt the Geneva model \citep{Eggenberger2008Ap&SS.316...43E} in which angular momentum is mainly transported by meridional currents \citep[see also][]{Belczynski2020ApJ...905L..15B},
$|\overrightarrow{\chi_1}|$ can increase from $\sim 0$ to $\sim 0.25$. As the spin of the merger product is dominated by the contribution of the more massive BH1,
the estimated
$\chi_{\rm eff}$ changes to be $-0.3 \sim 0.32$. We also note that, by using the Tayler-Spruit magnetic dynamo angular transport, \cite{Belczynski2020ApJ...905L..15B} inferred the natal spins of BBHs ($m_1$ = 84.9 $\msun$, $m_2$ = 64.6 $\msun$) that would merge within Hubble time to be $|\overrightarrow{\chi_1}| = 0.052$ and $|\overrightarrow{\chi_2}| = 0.523$. 

\section{Discussion}\label{sec:dis}

The natal kick plays a key role in the life of a compact star binary, as it affects not only the orbital parameters and systemic velocity, but also the binary evolutionary path \citep{Brandt1995MNRAS.274..461B}.
There is a general consensus that NSs are usually born with large kick velocities $v_{\rm kick,NS}\sim 200-500\,\rm kms^{-1}$ \citep{Lyne1994Natur.369..127L}. However, the origin of the SN kick is in debate. One possible mechanism is the asymmetric material ejection during the SN explosion, triggered by the large-scale hydrodynamic perturbation or convection instabilities in the SN core
\citep{Burrows1996PhRvL..76..352B, Goldreich1997upa..conf..269G, Scheck2004PhRvL..92a1103S, Nordhaus2012MNRAS.423.1805N, Gessner2018ApJ...865...61G}. Other investigations suggest that it may be related to the anisotropic neutrino emission from the proto-NS induced by strong magnetic filed \citep{Kusenko1996PhRvL..77.4872K, Lai1998ApJ...495L.103L, Maruyama2011PhRvD..83h1303M}. Besides, the topological current may be responsible for the natal kick \citep{Charbonneau2010JCAP...08..010C}.

The kick velocity distribution is also in active study.
\cite{Arzoumanian2002ApJ...568..289A} studied the velocity distribution of radio pulsars based on large-scale 0.4 GHz pulsar surveys,  and found a two-component velocity distribution with characteristic velocities of 90 and 500 kms$^{-1}$.
\cite{Hobbs2005MNRAS.360..974H} analyzed a catalogue of 233 pulsars with proper motion measurements, and suggested the NS natal kick distribution with a Maxwellian one-dimensional dispersion $\sigma_{\rm NS}=265\,\rm kms^{-1}$, which  is widely used in later studies. From the analysis of the proper motions of 28 pulsars using very long baseline array interferometry data, \cite{Verbunt2017A&A...608A..57V}  showed that a distribution with two Maxwellians improves significantly on a single Maxwellian for the young pulsar velocities.

Whether stellar-mass BHs receive such large kicks is also a matter of debate. A growing number of studies have been devoted to investigate the natal kicks of BHs relying on a variety of methods and data sets, such as the study of massive runaway and walkaway stars \citep{Blaauw1961BAN....15..265B, Donder1997A&A...318..812D, Renzo2019A&A...624A..66R, Aghakhanloo2022MNRAS.516.2142A}, BH X-ray binaries \citep{Mirabel2001Natur.413..139M,Jonker2004MNRAS.354..355J,Repetto2012MNRAS.425.2799R, Wong2012ApJ...747..111W, Wong2014ApJ...790..119W, Atri2019MNRAS.489.3116A, Kimball2022arXiv221102158K}, astrometric microlensing \citep{Andrews2022ApJ...930..159A}, and merging BBH GW events \citep{Abbott2021PhRvX..11b1053A, Abbott2021arXiv211103606T}.

In light of the observational constraints on the NS/BH natal kick velocities, several phenomenological and analytic kick prescriptions are proposed, mainly depending on the SN ejecta mass and remnant mass
\citep[e.g.,][]{Bray2018MNRAS.480.5657B, Giacobbo2020ApJ...891..141G, Mandel2020MNRAS.499.3214M, Richards2022arXiv220802407R}. Because the kick-induced orbital eccentricity determines the time-scale over which BBHs are expected to merger via GW radiation, the merging history of BBHs provide a probe to the natal kick received by BHs. Based on the premise that GW190521 is an IMRI with component masses of $ \sim170\,\msun $ and $ \sim 16\,\msun $ \citep{Nitz2021ApJ...907L...9N}, we examine the isolated binary evolution channel with three kick prescriptions. In the $ k1 $ and $ k2 $ prescriptions the BH natal kick is determined by the fallback fraction $f_{\rm fb}$, so massive BH experienced totally fallback ($f_{\rm fb}=1.0$) would receive no kick, while in $k3$ prescription BHs always receive a kick produced through asymmetric neutrino emission.

Our calculations indicate that, to produce the merger event, the less massive BH should receive a natal kick with velocity of a few hundred $\rm kms^{-1}$, thus preferring the $k3$ prescription. This is of particular interest since in most cases both BHs formed through totally fallback, and the conclusion is not sensitive to the choice of the CE efficiency $\alpha_{\rm CE}$.

We predict the merger rate density of GW190521-like systems $\mathcal{R}(z\leq1.1)\sim 4\times 10^{-5}-5\times 10^{-2} \,\rm Gpc^{-3}yr^{-1}$ if the BH natal kick is weighted to follow a Maxwell distribution of the NS kick with $\sigma_{\rm NS}=265$ kms$^{-1}$.  Under the interpretation that GW190521 is an almost equal mass ratio system, LIGO/Virgo collaboration reported the merger rate density of GW190521-like systems to be $0.13_{-0.11}^{+0.30}\,\rm Gpc^{-3} yr^{-1}$ with the effective spin parameter $\chi_{\rm eff}=0.08_{-0.36}^{+0.27}$ \citep{Abbott2020PhRvL.125j1102A, Abbott2020ApJ...900L..13A}. By employing a new estimate of the PPISN mass loss, \cite{Belczynski2020ApJ...905L..15B} obtained a merger rate density of $\sim 0.04\,\rm Gpc^{-3} yr^{-1}$ for such events via isolated binary evolution. \cite{Tanikawa2021ApJ...910...30T} estimated the merger rate density of Pop III BBHs (with total mass $\sim 130-260\,\msun$ and composing at least one $130-200\,\msun$ IMBH) about 0.01\,$\rm Gpc^{-3}yr^{-1}$. 
\cite{Hijikawa2022arXiv221107496H} performed a BPS calculation for very massive Population III stars and derived the property of the BBH mergers, adopting constant values for $\alpha_{\rm CE}\lambda$ in their CE evolution. In their `low mass + high mass' model, the resultant compact binaries consist of a stellar mass BH (below the PISN mass gap) and an IMBH (above the PISN mass gap) with mass ratio ranging from 0.15 to 0.35. The predicted merger rate density peaks at $z\sim 10$ with a value of $(1-10)\,\rm Gpc^{-3}yr^{-1}$, and declines to nearly zero at $z\leq3$ because of the very short delay time (less than 10 $\rm Myr$). In our Population II evolution channel, the merger rate peaks at $z\sim 2$ with the delay time ranging from $\sim 1.4-12.1\,\rm Gyr$.

\section{Summary}\label{sec:summary}
The third observing run operated by aLIGO and advanced Virgo discovered a massive BBH merger event GW190521, with a remnant total mass of $150_{{-17}}^{+29}\,\msun$, falling in the IMBH regime \citep{Abbott2020PhRvL.125j1102A}, and the component masses were estimated to be $(m1, m2) = (85_{{-14}}^{+21}$, $66_{{-18}}^{+17})\,\msun$ within 90\% credible region \citep[see also][]{ Barrera2022ApJ...929L...1B,Gamba2021arXiv210605575G}.
\cite{Nitz2021ApJ...907L...9N}, however, showed that GW190521 may be alternatively an IMRI, with the component masses of $m_1\sim 170\;\msun$ and $m_2\sim 16 \;\msun$, which happen to straddle the PISN mass gap. In the most recent analysis, \citet{Gamba2022NatAs.tmp..247G} revealed the BH masses to be $81_{{-25}}^{+62}\,\msun$ and $52_{{-32}}^{+32}\,\msun$  under the hypothesis that it was generated by the merger of two non-spinning BHs on hyperbolic orbits. So the nature of GW190521 is still uncertain.

Assuming the configuration of \cite{Nitz2021ApJ...907L...9N} for GW190521 (or similar systems to be discovered in the future), we perform BPS simulation to interpret the formation of GW190521-like systems via isolated binary evolution channel. Our analyses prefer that this merger event had evolved from primordial binary systems in metal-poor environment with $Z\leq 0.0016$. The majority of them are formed via an initial phase of stable RLOF before the formation of the BH1, followed by a CE phase triggered by collision at periastron when BH1's companion is a giant-like star in a close eccentric orbit. The initial ZAMS progenitor masses are expected to be $\mdi\sim 300-800\,\rm \msun$ and $\mci\sim 20-60\,\msun$, respectively, which are metallicity dependent.
By using the fallback-independent kick prescription, the merger event requires the primary and secondary BHs to receive natal kicks with velocities $v_{\rm kick,1}<50\,\rm kms^{-1}$ and $v_{\rm kick,2} <700\,\rm kms^{-1}$.
Our results support the hypothesis that BHs formed by direct core collapse receive considerable large natal kick. The predicted merger rate density for GW190521-like systems is $4\times 10^{-5}-5\times 10^{-2} \,\rm Gpc^{-3}yr^{-1}$ at $\zm \leq 1.1$. We also find that, using the traditional treatment of tidal interaction results in very small effective spin parameter, but if using the Geneva model instead, $\chi_{\rm eff}$ ranges from $-0.3$ to 0.32, roughly located within the interval $-0.51^{-0.11}_{+0.24}$ estimated by \cite{Nitz2021ApJ...907L...9N}.

\section*{Acknowledgments}
We thank the anonymous referee for their useful comments, which helped improve the manuscript. We are also grateful to Shi-Jie Gao for essential help with the calculation of $\lambda$. This work was supported by the Natural Science Foundation of China under grant No. 12041301 and 12121003, and the National Key Research and Development Program of China (2021YFA0718500).
We acknowledge use of the following PYTHON packages: ASTROPY \citep{Astropy2013A&A...558A..33A}, MATPLOTLIB \citep{Hunter2007CSE.....9...90H}, NUMPY \citep{van2011CSE....13b..22V} and SciPY \citep{Virtanen2020NatMe..17..261V}. 

\section*{Data Availability}
All data underlying this article will be shared on reasonable request to the corresponding authors.

\begin{appendix}
\section{THe CE parameter $\lambda$ for very massive stars}\label{app}
In our stellar models, we neglect any stellar rotation. For stellar winds, we use the  {\tt\string Vink et al.\,wind} prescription as mentioned in Section \ref{sub:wind}, but the result in \cite{Nugis2000A&A...360..227N} for Wolf-Rayet stars. The convective regions are determined by the Schwarzschild criterion using $\alpha_{\rm MLT}=1.73$ \citep{MacLeod2017ApJ...838...56M}. And we assume an exponential overshoot \citep{Herwig2000A&A...360..952H} by 1\% of the scale height.

\begin{figure*}
	\centering
	\includegraphics[width=1.10\linewidth]{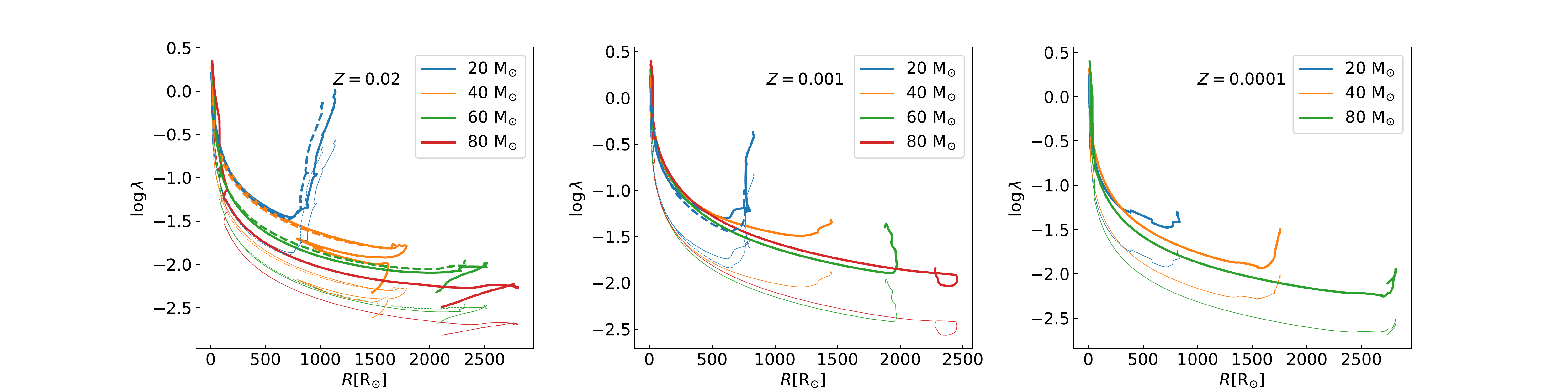}
	\caption{Evolution of binding energy parameter $\lambda_{\rm b}$ (the thick solid lines) and $\lambda_{\rm g}$ (the thin solid lines) vary with stellar radius for different massive stars. Results with $Z =0.02$ (the left panel), $Z=0.001$ (the middle panel) and $Z=0.0001$ (the right panel). Models with $M = 20,40,60\msun$ at $Z=0.02$ in \protect\cite{Wang2016RAA....16..126W} and $M = 20\msun$ at $Z=0.001$ in \protect\cite{Xu2010ApJ...716..114X} are also shown with dashed lines as controls.}
	\label{fig:lbd_com}
\end{figure*}

\begin{figure*}
	\centering
	\includegraphics[width=1.10\linewidth]{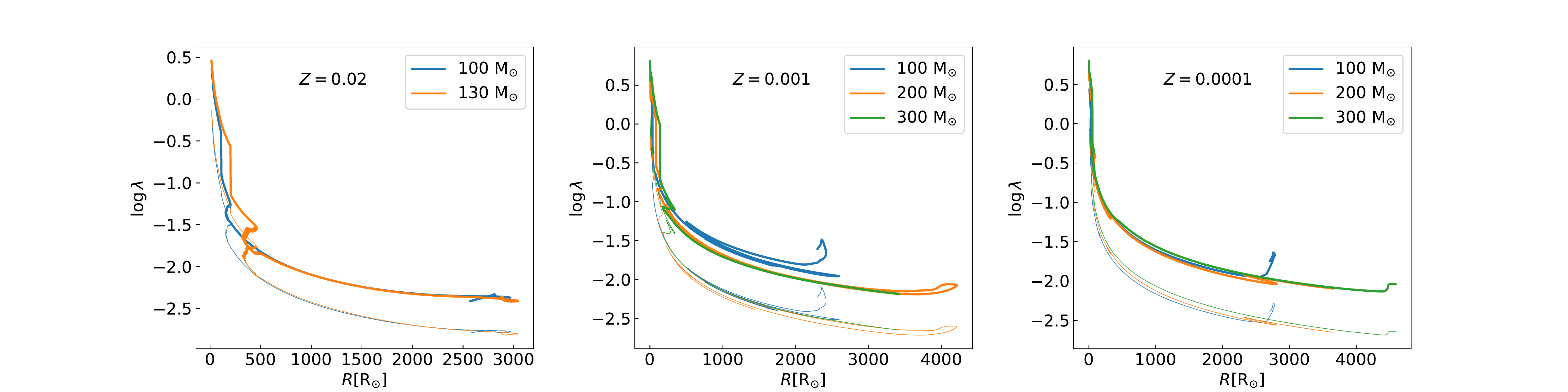}
	\caption{Same as Fig. \ref{fig:lbd_com} but for more massive stars.}
	\label{fig:lbd_all}
\end{figure*}

Fig. \ref{fig:lbd_com} and Fig. \ref{fig:lbd_all} show the evolution of $\lambda_{\rm b}$(the thick solid lines) and $\lambda_{\rm g}$(the thin solid lines) with stellar radius for different masses at $Z$ = $0.02$, $0.001$ and $0.0001$.
We make polynomial fitting of the results with
\begin{equation*}
	\log(\lambda) =\sum_{i=0}^{10}n_{i}(R/\rsun)^i
\end{equation*}
where $R$ the stellar radius. Table \ref{fitting} and \ref{fitting2} provide the coefficients $n_{i(i = 0, 10)}$ for stars with selected masses. The goodness of fit $\chi^2\simeq 0.998$.

\clearpage

		\begin{sidewaystable*}[!htbp]
			\vspace{17cm}
		\caption{Fitting coefficients for $\lambda_{\rm b}$.}
		\begin{tabular}{ccccccccccccc}
			\hline\hline
			$\lambda$&$Mass[\msun]$&$n_0$& $n_1$& $n_2$& $n_3$& $n_4$& $n_5$& $n_6$& $n_7$& $n_8$& $n_9$& $n_{10}$\\
			\hline
			$Z$=0.02 &&&&&&&&&&&&\\
			&40&0.5435& - 0.03184&0.0002715&- 1.227e-06&3.209e-09&- 5.196e-12& 5.375e-15& - 3.56e-18&1.461e-21& - 3.382e-25&3.378e-29\\
			&60& 0.6051&- 0.02582& 0.0001443&- 4.475e-07& 8.257e-10&- 9.575e-13&7.162e-16&- 3.451e-19&1.034e-22&- 1.754e-26&1.286e-30\\
			&80&0.6566&- 0.02267& 9.957e-05&- 2.494e-07&3.81e-10&- 3.721e-13&2.367e-16 & - 9.745e-20&2.498e-23&- 3.616e-27& 2.254e-31\\
			&100& 0.6693&- 0.01809&5.575e-05& - 9.112e-08&7.847e-11&- 2.704e-14&- 9.748e-18&1.372e-20&- 5.748e-24&1.129e-27&-8.781e-32\\
			&130& 0.5937&- 0.01164&3.108e-05&- 6.794e-08&1.096e-10&- 1.178e-13&8.156e-17&- 3.571e-20&9.536e-24&- 1.416e-27&8.955e-32\\
			\hline
			$Z$=0.001 &&&&&&&&&&&&\\
			&40& 0.5478&- 0.0358&0.0004194&- 2.648e-06&9.582e-09&- 2.114e-11&2.94e-14&- 2.588e-17&1.398e-20&- 4.228e-24&5.482e-28\\
			&60& 0.6674&- 0.03372&0.0002993&- 1.399e-06& 3.726e-09&- 6.048e-12& 6.193e-15 &- 4.02e-18& 1.604e-21& - 3.588e-25&3.446e-29\\
			&80& 0.7491&- 0.03138&0.0002319&- 8.724e-07&1.845e-09&- 2.363e-12&1.906e-15&- 9.73e-19&3.053e-22& - 5.37e-26&4.055e-30\\
			&100&0.7908&- 0.02809& 0.0001731&- 5.611e-07& 1.05e-09&- 1.213e-12&8.937e-16&- 4.206e-19&1.225e-22&- 2.009e-26&1.421e-30 \\
			&150&0.6241 &- 0.009325 &- 1.416e-05 &1.386e-07 &- 2.997e-10 &3.234e-13 &- 2.025e-16 & 7.697e-20 &- 1.753e-23 &2.203e-27&-1.175e-31 \\	
			&200& 0.7362&- 0.01171&2.096e-05&- 1.864e-08&9.033e-12&- 2.438e-15& 3.444e-19& -1.986e-23\\
			&250&0.7734 &- 0.01013 &1.674e-05 &- 8.327e-09 &- 8.46e-12 &1.412e-14 &- 8.661e-18 &2.889e-21 &- 5.523e-25 &5.694e-29 &-2.46e-33 \\
			&300&0.8114&- 0.009691&1.45e-05&- 1.085e-08&4.247e-12 &- 8.348e-16&6.517e-20\\
			\hline
			$Z$=0.0001 &&&&&&&&&&&&\\
			&40&0.3171& - 0.01658&6.395e-05& - 1.197e-07&1.126e-10&- 5.171e-14&9.225e-18\\
			&60& 0.4729&- 0.02887& 0.0002611&- 1.259e-06&3.443e-09&- 5.681e-12&5.86e-15&- 3.803e-18 & 1.508e-21&- 3.338e-25&3.161e-29\\
			&80&  0.7627&- 0.03771&0.000268&- 9.484e-07& 1.87e-09&- 2.22e-12&1.649e-15&- 7.712e-19&2.207e-22& - 3.528e-26&	2.414e-30\\
			&100& 0.6976&- 0.02694& 0.0001746& - 5.963e-07&1.16e-09&- 1.371e-12&1.02e-15& - 4.795e-19 &1.382e-22&- 2.229e-26&	1.54e-30 \\
			&150& 0.6742&- 0.01649 &5.084e-05 &- 7.856e-08 &6.492e-11 &- 2.924e-14 &6.767e-18 & -6.291e-22\\
			&200&0.7422& - 0.01526&4.124e-05& - 4.653e-08& 7.804e-12& 3.139e-14 &- 3.343e-17&1.61e-20&- 4.216e-24& 5.817e-28&-3.318e-32\\
			&250&0.516 &- 0.002953 &1.331e-06 &- 2.585e-10 &1.748e-14\\
			&300& 0.8598&- 0.0158&4.625e-05&- 6.992e-08&5.992e-11&- 3.098e-14&9.893e-18&- 1.925e-21&2.14e-25&  - 1.141e-29&1.573e-34\\
			\hline\hline
		\end{tabular}
		\label{fitting}
		\end{sidewaystable*}

		\begin{sidewaystable*}[!htbp]
			\vspace{17cm}
		\caption{Same as Table \ref{fitting} but for $\lambda_{\rm g}$.}
		\begin{tabular}{ccccccccccccc}
			\hline\hline
			$\lambda$&$Mass[\msun]$&$n_0$& $n_1$& $n_2$& $n_3$& $n_4$& $n_5$& $n_6$& $n_7$& $n_8$& $n_9$& $n_{10}$\\
			\hline
			$Z$=0.02 &&&&&&&&&&&&\\
			&40&0.144& - 0.03683&0.0003541&- 1.767e-06&5.024e-09 &- 8.734e-12& 9.615e-15&- 6.731e-18&2.903e-21& - 7.036e-25&7.33e-29\\
			&60&0.03541& - 0.02224&0.0001293&- 4.263e-07&8.269e-10 & - 9.924e-13& 7.581e-16&- 3.693e-19&1.111e-22&- 1.882e-26 &1.373e-30\\
			&80& - 0.05606&- 0.01411&3.883e-05&- 4.339e-08&- 1.224e-11&8.372e-14 & - 9.515e-17& 5.486e-20&- 1.774e-23&3.062e-27&-2.203e-31\\
			&100&0.03445& - 0.01607&  5.61e-05& - 1.16e-07&1.474e-10 &- 1.183e-13&6.02e-17&- 1.896e-20 & 3.461e-24&- 3.122e-28& 8.471e-33 \\
			&130&0.01228&- 0.01478 &7.119e-05&- 2.132e-07&3.702e-10& - 3.893e-13&2.564e-16 &- 1.065e-19&2.712e-23&- 3.869e-27&2.368e-31 \\
			\hline
			$Z$=0.001 &&&&&&&&&&&&\\
			&40&0.07005& - 0.03453& 0.0003868 & - 2.385e-06&  8.509e-09& - 1.861e-11&  2.574e-14&- 2.258e-17& 1.217e-20& - 3.675e-24&	4.76e-28\\
			&60&0.08348& - 0.02932& 0.0002426& - 1.088e-06& 2.807e-09 &- 4.432e-12&  4.425e-15&  - 2.802e-18& 1.091e-21& - 2.382e-25 &2.231e-29\\
			&80& 0.09617& - 0.02641& 0.0001841 &- 6.737e-07 & 1.4e-09& - 1.769e-12 &  1.41e-15& - 7.117e-19& 2.208e-22&- 3.839e-26&2.864e-30\\
			&100&0.103& - 0.02412&  0.0001445&- 4.624e-07&8.514e-10 &- 9.62e-13& 6.891e-16 &- 3.14e-19&   8.821e-23& - 1.393e-26&9.459e-31\\
			&150& 0.1182&- 0.02012& 0.0001171&- 3.371e-07& 5.206e-10& - 4.745e-13& 2.686e-16& - 9.566e-20&2.088e-23&  - 2.555e-27&1.342e-31  \\
			&200&0.1519&  - 0.0191 & 9.396e-05&- 2.224e-07 & 2.831e-10 &- 2.134e-13&1.003e-16 & - 2.972e-20& 5.408e-24 &- 5.52e-28 &2.422e-32\\
			&250& 0.1054& - 0.01521& 6.26e-05&- 1.315e-07 &1.512e-10 &- 1.037e-13 &4.446e-17 &- 1.203e-20&1.999e-24 &- 1.863e-28 &7.466e-33\\
			&300&0.1743& - 0.01709& 8.94e-05 & - 2.608e-07& 4.266e-10 &- 4.174e-13&2.544e-16 &- 9.753e-20& 2.287e-23&- 2.997e-27&1.683e-31\\
			\hline
			$Z$=0.0001 &&&&&&&&&&&&\\
			&40&- 0.1711& - 0.01538&5.801e-05&- 1.069e-07&9.934e-11&- 4.504e-14&7.935e-18\\
			&60&- 0.04446& - 0.02763&0.0002392&- 1.129e-06&3.042e-09 &- 4.968e-12& 5.082e-15 &- 3.275e-18&  1.291e-21&  - 2.845e-25&2.684e-29\\
			&80& 0.04396& - 0.02577& 0.0001593&- 5.317e-07&1.025e-09&- 1.21e-12&9.021e-16& - 4.257e-19&1.232e-22 & - 1.992e-26& 1.379e-30\\
			&100&- 0.0111&- 0.02101& 0.0001237&- 4.057e-07&7.76e-10&- 9.134e-13& 6.808e-16&- 3.218e-19& 9.345e-23&- 1.52e-26& 	1.059e-30 \\
			&150&0.0299 &- 0.01725 &8.402e-05 &- 2.49e-07 &4.516e-10 &- 5.153e-13 & 3.758e-16 &- 1.745e-19 &4.979e-23 &- 7.951e-27 &5.435e-31\\
			&200& 0.03978&  - 0.01477& 5.592e-05& - 1.205e-07&  1.543e-10&- 1.235e-13& 6.328e-17&- 2.079e-20&   4.236e-24 & - 4.876e-28& 	2.426e-32\\
			&250&- 0.08852 &- 0.008596 &1.748e-05 &- 1.76e-08 &9.417e-12 &- 2.746e-15 &4.127e-19 &-2.503e-23 \\
			&300&0.07319& - 0.01425& 4.956e-05& - 9.459e-08&  1.05e-10&- 7.172e-14& 3.104e-17 &- 8.531e-21& 1.441e-24&  - 1.365e-28&	5.542e-33\\
			\hline\hline
		\end{tabular}
		\label{fitting2}
		\end{sidewaystable*}

\end{appendix}



\label{lastpage}

\end{document}